\def\verPreprint{1}
\def\verPAPER{2}
\def\ver{1}

\ifx\ver\verPreprint
  \documentclass[a4paper,english,11pt]{article}
  \usepackage[bindingoffset=0.5cm,left=2.5cm,right=2.5cm,top=2.5cm,bottom=2.5cm,footskip=1.0cm]{geometry}
\fi
\ifx\ver\verPAPER
  \documentclass[preprint,12pt]{elsarticle}
\fi

\usepackage{amsmath}
\usepackage{amssymb}
\usepackage{xspace}
\usepackage{float}
\usepackage{hyperref}
\usepackage{subcaption}
\usepackage{relsize}
\usepackage{multirow}
\usepackage{xcolor}
\usepackage{graphicx}
\usepackage{lineno}
\usepackage[english]{babel}
\usepackage[bb=dsserif]{mathalpha}
\usepackage{bbding}

\ifx\ver\verPreprint
  \usepackage{authblk}
\fi

\ifx\ver\verPAPER
  \modulolinenumbers[5]
\fi

\newcommand{\Pagne}{\ensuremath{\bar{\nu}_{\textrm{e}}}\xspace}
\newcommand{\Pagngm}{\ensuremath{\bar{\nu}_{\mu}}\xspace}
\newcommand{\Pagngt}{\ensuremath{\bar{\nu}_{\tau}}\xspace}
\newcommand{\Paq}{\ensuremath{\bar{\textrm{q}}}\xspace}
\newcommand{\Paqb}{\ensuremath{\bar{\textrm{b}}}\xspace}
\newcommand{\Paqc}{\ensuremath{\bar{\textrm{c}}}\xspace}
\newcommand{\PcB}{\ensuremath{\textrm{B}_{\textrm{c}}^{+}}\xspace}
\newcommand{\Pe}{\ensuremath{\textrm{e}}\xspace}
\newcommand{\Pem}{\ensuremath{\textrm{e}^{-}}\xspace}
\newcommand{\Pep}{\ensuremath{\textrm{e}^{+}}\xspace}
\newcommand{\Pgg}{\ensuremath{\gamma}\xspace}
\newcommand{\Pggx}{\ensuremath{\gamma^{\ast}}\xspace}
\newcommand{\Pgm}{\ensuremath{\mu}\xspace}
\newcommand{\Pgmm}{\ensuremath{\mu}^{-}\xspace}
\newcommand{\Pgmp}{\ensuremath{\mu}^{+}\xspace}
\newcommand{\Pgn}{\ensuremath{\nu}\xspace}
\newcommand{\Pgne}{\ensuremath{\nu_{\textrm{e}}}\xspace}
\newcommand{\Pgngm}{\ensuremath{\nu_{\mu}}\xspace}
\newcommand{\Pgngt}{\ensuremath{\nu_{\tau}}\xspace}
\newcommand{\Pgpm}{\ensuremath{\pi^{-}}\xspace}
\newcommand{\Pgpp}{\ensuremath{\pi^{+}}\xspace}
\newcommand{\Pgppm}{\ensuremath{\pi^{\pm}}\xspace}
\newcommand{\Pgpz}{\ensuremath{\pi^{\textrm{0}}}\xspace}
\newcommand{\Pgt}{\ensuremath{\tau}\xspace}
\newcommand{\Pgtm}{\ensuremath{\tau^{-}}\xspace}
\newcommand{\Pgtp}{\ensuremath{\tau^{+}}\xspace}
\newcommand{\Pgtpm}{\ensuremath{\tau^{\pm}}\xspace}
\newcommand{\PH}{\ensuremath{\textrm{H}}\xspace}
\newcommand{\PHm}{\ensuremath{\textrm{H}^{-}}\xspace}
\newcommand{\PHp}{\ensuremath{\textrm{H}^{+}}\xspace}
\newcommand{\PHpm}{\ensuremath{\textrm{H}^{\pm}}\xspace}

\newcommand{\PKm}{\ensuremath{\textrm{K}^{-}}\xspace}
\newcommand{\PKp}{\ensuremath{\textrm{K}^{+}}\xspace}
\newcommand{\PKpm}{\ensuremath{\textrm{K}^{\pm}}\xspace}
\newcommand{\PKz}{\ensuremath{\textrm{K}^{\textrm{0}}}\xspace}
\newcommand{\PKzL}{\ensuremath{\textrm{K}_{\textrm{L}}^{\textrm{0}}}\xspace}
\newcommand{\PKzS}{\ensuremath{\textrm{K}_{\textrm{S}}^{\textrm{0}}}\xspace}
\newcommand{\Pl}{\ensuremath{\ell}\xspace}
\newcommand{\Pp}{\ensuremath{\textrm{p}}\xspace}
\newcommand{\Pq}{\ensuremath{\textrm{q}}\xspace}
\newcommand{\Pqb}{\ensuremath{\textrm{b}}\xspace}
\newcommand{\Pqc}{\ensuremath{\textrm{c}}\xspace}

\newcommand{\PWm}{\ensuremath{\textrm{W}^{-}}\xspace}
\newcommand{\PWp}{\ensuremath{\textrm{W}^{+}}\xspace}

\newcommand{\PZ}{\ensuremath{\textrm{Z}}\xspace}

\newcommand{\tauh}{\ensuremath{\Pgt_{\textrm{h}}}\xspace}
\newcommand{\h}{\ensuremath{\textrm{h}}\xspace}
\newcommand{\hpm}{\ensuremath{\textrm{h}^{\pm}}\xspace}
\newcommand{\hplus}{\ensuremath{\textrm{h}^{+}}\xspace}
\newcommand{\hminus}{\ensuremath{\textrm{h}^{-}}\xspace}
\newcommand{\hzero}{\ensuremath{\textrm{h}^{\textrm{0}}}\xspace}

\newcommand{\pT}{\ensuremath{p_{\textrm{T}}}\xspace}

\newcommand{\kt}{\ensuremath{k_{\textrm{t}}}\xspace}

\newcommand{\GeV}{\ensuremath{\textrm{GeV}}\xspace}
\newcommand{\TeV}{\ensuremath{\textrm{TeV}}\xspace}

\newcommand{\gen}{\ensuremath{\textrm{gen}}\xspace}
\newcommand{\rec}{\ensuremath{\textrm{rec}}\xspace}
\newcommand{\jet}{\ensuremath{\textrm{jet}}\xspace}

\newcommand{\jetR}{\ensuremath{\langle r \rangle}\xspace}
\newcommand{\jetM}{\ensuremath{M_{\jet}}\xspace}
\newcommand{\Dtau}{\ensuremath{\mathcal{D}_{\Pgt}}\xspace}
\newcommand{\R}{\ensuremath{\rm I\!R}\xspace}
\newcommand{\degr}{\ensuremath{^{\circ}}\xspace}
\newcommand{\iso}{\ensuremath{\textrm{iso}}\xspace}
\newcommand{\misid}{\ensuremath{\textrm{misid}}\xspace}

\newcommand{\conjunction}{\ensuremath{\,\, \& \,\,}\xspace}
\newcommand{\cf}{cf.\xspace}
\newcommand{\eg}{e.g.\xspace}
\newcommand{\efficiency}{\ensuremath{\varepsilon_{\Pgt}}\xspace}
\newcommand{\fakerate}{\ensuremath{P_{\textrm{misid}}}\xspace}
\newcommand{\X}{\ensuremath{\textrm{X}}\xspace}
\newcommand{\genX}{\ensuremath{\textrm{\gen-X}}\xspace}
\newcommand{\dxy}{\ensuremath{d_{\textrm{xy}}}\xspace}
\newcommand{\sigmaxy}{\ensuremath{\sigma_{d{\textrm{xy}}}}\xspace}
\newcommand{\dz}{\ensuremath{d_{\textrm{z}}}\xspace}
\newcommand{\sigmaz}{\ensuremath{\sigma_{d{\textrm{z}}}}\xspace}
\newcommand{\indicator}{\ensuremath{\mathcal{O}}\xspace}
\newcommand{\ch}{\ensuremath{\textrm{ch}}\xspace}
\newcommand{\nh}{\ensuremath{\textrm{nh}}\xspace}
\newcommand{\ig}{\ensuremath{\textrm{i}}\xspace}
\newcommand{\og}{\ensuremath{\textrm{o}}\xspace}

\ifx\ver\verPreprint
  \title{Tau lepton identification and reconstruction: a new frontier for jet-tagging ML algorithms}
  \author[1]{Torben Lange}
  \author[1]{Saswati Nandan}
  \author[1]{Joosep Pata}
  \author[1]{Laurits Tani}
  \author[1]{Christian Veelken}
  \affil[1]{National Institute of Chemical Physics and Biophysics (NICPB), R\"{a}vala pst 10, 10143 Tallinn, Estonia}
\fi

\begin{document}

\ifx\ver\verPAPER
\begin{frontmatter}
    \journal{Computer Physics Communications}
    \title{Tau lepton identification and reconstruction: a new frontier for jet-tagging ML algorithms}
    \author[a]{Torben Lange}
    \author[a]{Saswati Nandan}
    \author[a]{Joosep Pata}
    \author[a]{Laurits Tani}
    \author[a]{Christian Veelken\corref{author}}
    \cortext[author]{Corresponding author.\\\textit{E-mail address:} christian.veelken@cern.ch}
    \address[a]{National Institute of Chemical Physics and Biophysics (NICPB), R\"{a}vala pst 10, 10143 Tallinn, Estonia}

    \begin{abstract}
      Identifying and reconstructing hadronic $\Pgt$ decays ($\tauh$) is an important task at current and future high-energy physics experiments,
      as $\tauh$ represent an important tool to analyze the production of Higgs and electroweak bosons as well as to search for physics beyond the Standard Model. 
      The identification of $\tauh$ can be viewed as a generalization and extension of jet-flavour tagging, which has in the recent years undergone significant progress
      due to the use of deep learning.
      Based on a granular simulation with realistic detector effects and a particle flow-based event reconstruction,
      we show in this paper that deep learning-based jet-flavour-tagging algorithms are powerful $\tauh$ identifiers.
      Specifically, we show that jet-flavour-tagging algorithms such as LorentzNet and ParticleTransformer can be adapted in an end-to-end fashion 
      for discriminating $\tauh$ from quark and gluon jets.
      We find that the end-to-end transformer-based approach significantly outperforms contemporary state-of-the-art $\tauh$ reconstruction and identification algorithms
      currently in use at the Large Hadron Collider.
    \end{abstract}

    \begin{keyword}
      tau identification; machine learning; ParticleTransformer; LorentzNet; DeepTau
    \end{keyword}
\end{frontmatter}
\fi

\ifx\ver\verPreprint
  %\linenumbers
  \maketitle
  \begin{abstract}
    Identifying and reconstructing hadronic $\Pgt$ decays ($\tauh$) is an important task at current and future high-energy physics experiments,
    as $\tauh$ represent an important tool to analyze the production of Higgs and electroweak bosons as well as to search for physics beyond the Standard Model. 
    The identification of $\tauh$ can be viewed as a generalization and extension of jet-flavour tagging, which has in the recent years undergone significant progress
    due to the use of deep learning.
    Based on a granular simulation with realistic detector effects and a particle flow-based event reconstruction,
    we show in this paper that deep learning-based jet-flavour-tagging algorithms are powerful $\tauh$ identifiers.
    Specifically, we show that jet-flavour-tagging algorithms such as LorentzNet and ParticleTransformer can be adapted in an end-to-end fashion 
    for discriminating $\tauh$ from quark and gluon jets.
    We find that the end-to-end transformer-based approach significantly outperforms contemporary state-of-the-art $\tauh$ reconstruction and identification algorithms
    currently in use at the Large Hadron Collider.
  \end{abstract}
\fi

\section{Introduction}

Jets constitute an important experimental signature at current and future high-energy physics experiments.
The task of identifying or ``tagging'' the parton that originated the jet based on its constituent structure has undergone remarkable progress in recent years, 
driven by supervised machine learning on open datasets, see \eg Refs.~\cite{Larkoski:2017jix,Kogler:2018hem} for a review.
In particular, the application of advanced deep-learning (DL) techniques, originally developed for image, natural-language and point-cloud processing, 
has enabled fine-grained and robust classification based on the rich information available in the particle constituents of a jet.
Hadronic $\Pgt$ decays ($\tauh$) can be regarded as a special type of highly-collimated jets of low particle multiplicity.
This motivated us to study the prospects for applying the same techniques to the task of identifying $\tauh$ based on the particle content of the $\tauh$ candidates.
The results of this study are presented in this paper.

With a lifetime of $2.9 \times 10^{-13}$ seconds, the $\Pgt$ lepton decays almost instantaneously. It can thus not be detected directly, but the particles produced in the $\Pgt$ decay can.
The reconstruction and identification of $\Pgt$ leptons is thus based on the reconstruction of the $\Pgt$ decay products.
In about one-third of the cases the $\Pgt$ decays to an electron or muon plus two neutrinos.
In the remaining two-thirds of the cases, the $\Pgt$ decays into one neutrino plus a system of hadrons, consisting of typically either one or three charged pions ($\Pgppm$) or kaons ($\PKpm$) and up to two neutral pions ($\Pgpz$). The $\Pgpz$ mesons decay nearly instantly and almost exclusively to a pair of photons. Decays of $\Pgt$ leptons into five charged mesons are rare~\cite{ParticleDataGroup:2022pth}.
In the energy range of interest, the $\Pgppm$ and $\PKpm$ produced in the $\Pgt$ decays are difficult to distinguish experimentally. We collectively refer to them using the symbol $\hpm$.
We further introduce the symbol $\tauh$ to refer to the system of all hadrons produced in the $\Pgt$ decay.
The decays of $\Pgt^{+}$ and $\Pgtm$ are related by charge conjugation invariance.

The $\Pgt$ lepton is instrumental for Standard Model (SM) precision measurements as well as for searches for physics beyond the SM (BSM).
Measurements of the $\Pgt$ lepton's properties, such as its lifetime, mass, and branching fractions allow to test the universality between lepton generations of the charged-current coupling,
while measurements of its spin polarization allow to probe the neutral-current coupling of the $\Pgt$ lepton~\cite{Dam:2021ibi,Pich:2020qna}.
Measurements of hadronic $\Pgt$ decays allow to study perturbative and non-perturbative effects in quantum chromodynamics.
A variety of models for BSM physics predict new particles that predominantly decay to $\Pgt$ leptons, such as models with extended gauge symmetries that manifest themselves through heavy charged and neutral gauge bosons~\cite{ATLAS:2017eiz,ATLAS:2018ihk,CMS:2022ncp}, models of third generation lepto-quarks~\cite{ATLAS:2019qpq}, supersymmetric models~\cite{ATLAS:2017qwn,ATLAS:2018xpg,ATLAS:2018zzq,CMS:2019eln,CMS:2019zmn,ATLAS:2019gti,ATLAS:2021pzz,CMS:2021cox,CMS:2023yzg}, and models with an extended Higgs sector~\cite{ATLAS:2018gfm,CMS:2019bfg,ATLAS:2020zms,CMS:2022goy,CMS:2021yci}. 
The $\Pgt$ lepton is also important for tests of lepton-flavour violation,
which may reveal itself in $\Pgt$ decays~\cite{Dam:2021ibi,Pich:2020qna} as well as in lepton-flavour violating decays of $\PZ$~\cite{Dam:2018rfz,ATLAS:2018sky} 
and Higgs ($\PH$)~\cite{Harnik:2012pb,CMS:2017con} bosons to a $\Pgt$ lepton and an electron or muon.
The sizable coupling of the $\Pgt$ lepton to the $\PH$ boson has been used to probe the $\PH$ boson coupling to fermions~\cite{ATLAS:2015xst,CMS:2014suk}
and to study the Higgs potential~\cite{ATLAS:2022xzm,CMS:2022hgz}.

In this paper, we focus on the identification of hadronic $\Pgt$ decays. The $\Pe$ and $\Pgm$ produced in $\Pgt$ decays can be reconstructed and identified using standard algorithms for electron and muon reconstruction\footnote{
The small distance that a $\Pgt$ lepton typically travels between its production and decay, results in a finite impact parameter of the electron or muon track with respect to the primary event vertex, which provides a handle to distinguish $\Pe$ and $\Pgm$ produced in the primary $\Pep\Pem$ collision from those resulting from $\Pgt$ decays.
The neutrinos produced in leptonic $\Pgt$ decays provide another handle to this end. Their momenta can be inferred from energy and momentum conservation or computed, with improved resolution, by means of dedicated algorithms~\cite{Elagin:2010aw,Bianchini:2016yrt}.}.
Our study is performed in electron--positron ($\Pep\Pem$) collisions at the CLIC linear collider~\cite{Linssen:2012hp} at a center-of-mass energy of $\sqrt{s} = 380$~\GeV.
The motivation for performing the study in $\Pep\Pem$ collisions is twofold:
first, because a detailed simulation of the CLICdet detector~\cite{CLICdp:2017vju}
and a performant event reconstruction based on the particle-flow (PF) approach~\cite{Marshall:2012hh,Marshall:2012ry,Marshall:2015rfa} is publicly available.
The PF approach combines information provided by tracking detectors with calorimeter information.
Our experience at the Large Hadron Collider (LHC) is that the PF approach greatly benefits the identification of hadronic $\Pgt$ decays.
Second, the literature on $\tauh$ identification at future high-energy $\Pep\Pem$ experiments is sparse and often based on simple algorithms
similar to those used by the ATLAS and CMS collaborations during the start-up of the LHC~\cite{CMS:2008dex,ATLAS:2010tba}.
Relevant references are~\cite{LCD-2010-009,CEPCStudyGroup:2018ghi,Giagu:2022gmq}.
Refs.~\cite{Behnke:2013lya,Tran:2015nxa,Xu:2017lgs,Dam:2021ibi} focus on distinguishing between individual hadronic decay modes of the $\Pgt$ lepton,
which is important in particular for measurements of its branching fractions and for $\Pgt$ spin polarization measurements.

The main result of this paper is that the advancements in DL techniques that drove the progress in jet-flavour tagging
significantly improve the identification of hadronic $\Pgt$ decays.
We compare the performance of two recently published algorithms for jet-flavour tagging,
LorentzNet~\cite{Gong:2022lye} and ParticleTransformer~\cite{Qu:2022mxj},
to state-of-the-art $\tauh$ reconstruction and identification algorithms currently in use at the LHC,
the ``hadrons-plus-strips'' (HPS)~\cite{CMS:2015pac,CMS:2018jrd} and DeepTau~\cite{CMS:2022prd} algorithms.
The reconstruction and identification of $\tauh$ by the latter algorithms proceeds in two steps:
In the first step the $\tauh$ are reconstructed by the HPS algorithm,
and in the second step the reconstructed $\tauh$ are identified by the DeepTau algorithm,
where ``identification'' refers to discriminating the $\tauh$ from quark and gluon jets.
The HPS algorithm has been developed by domain experts and does not employ machine-learning techniques,
while the DeepTau algorithm is based on a convolutional deep neural network (DNN).
We have retrained the DeepTau algorithm for $\Pep\Pem$ collisions, using the same event samples for its training as for the LorentzNet and ParticleTransformer algorithms.
We use the combination of the HPS algorithm for $\tauh$ reconstruction and DeepTau algorithm for $\tauh$ identification as reference for state-of-the-art algorithms
against which we compare the performance of the DL-based algorithms. The latter perform the tasks of $\tauh$ reconstruction and identification in a single step,
following an end-to-end approach.
Our choice of the LorentzNet and ParticleTransformer algorithms is based on Ref.~\cite{Gong:2022lye},
which reported that the LorentzNet algorithm outperforms alternative DL-based jet-flavour tagging algorithms
such as the ResNeXt-50~\cite{xie2017aggregated}, ParticleFlowNetwork~\cite{Komiske:2018cqr}, and ParticleNet~\cite{Qu:2019gqs} algorithms
on different jet tagging tasks.
The ParticleTransformer algorithm has been developed by the same group of authors as the LorentzNet algorithm.
It extends the latter by using additional input variables, which we expect may increase the $\tauh$ identification performance.
We believe this result to be applicable to future high-energy $\Pep\Pem$ experiments such as
CEPC~\cite{CEPCStudyGroup:2018ghi}, CLIC~\cite{Linssen:2012hp}, FCC-ee~\cite{FCC:2018evy}, and ILC~\cite{ILC:2013jhg}
as well as to proton--proton ($\Pp\Pp$) collisions at the LHC.

The paper is structured as follows: In Section~\ref{sec:MCSamples_and_event_reconstruction},
we detail the simulated samples of $\tauh$ and of quark and gluon jets that we use to study the $\tauh$ reconstruction and identification performance. The reconstruction of muons, electrons, photons, charged and neutral hadrons via the PF method, which are used as input for the $\tauh$ identification, is described in the same section.
In Section~\ref{sec:TauIDAlgos}, we present the LorentzNet and ParticleTransformer algorithms.
The performance of these algorithms is compared to the performance of the HPS and DeepTau algorithms in Section~\ref{sec:Results}.
The HPS and DeepTau algorithms are described in the appendix.
Our motivation for documenting the HPS and DeepTau algorithms in the appendix is to concisely summarize the relevant information of Refs.~\cite{CMS:2015pac,CMS:2018jrd,CMS:2022prd}
and to record the few adjustments that we have made to adapt the algorithms to $\Pep\Pem$ collisions.
We conclude the paper with a summary and an outlook in Section~\ref{sec:Summary}.

\section{Monte Carlo samples and event reconstruction}
\label{sec:MCSamples_and_event_reconstruction}

The optimization and subsequent performance evaluation of the $\tauh$ identification algorithms is based on a set of Monte Carlo (MC) event samples.
The samples are generated for $\Pep\Pem$ collisions at a center-of-mass energy of $\sqrt{s} = 380$~\GeV, using the program \textsc{Pythia8}~\cite{Bierlich:2022pfr}.
We generate 1 million ``signal'' events of $\PZ/\Pggx \to \Pgt\Pgt$ and $\PZ\PH$, $\PH \to \Pgt\Pgt$ each and 2 million ``background'' events of $\PZ/\Pggx \to \Pq\Paq'$.
The $\PZ\PH$, $\PH \to \Pgt\Pgt$ sample is used to train the LorentzNet, ParticleTransformer, and DeepTau algorithms and to evaluate their performance.
The $\PZ/\Pggx \to \Pgt\Pgt$ sample is used as a cross check when evaluating the algorithms' performance, to examine that the algorithms did not exploit differences in event kinematics
between the $\PZ\PH$, $\PH \to \Pgt\Pgt$ signal and the $\PZ/\Pggx \to \Pq\Paq'$ background,
as this would result in an overly optimistic assessment of the algorithms' performance.
To this end, suitable chosen weights are applied to the samples of $\tauh$ and jets used for the training of the DeepTau, LorentzNet, and ParticleTransformer algorithms.
The weights are chosen such that the distributions in polar angle $\theta$ and transverse momentum $\pT$ of the reconstructed jets become identical 
for the $\PZ\PH$, $\PH \to \Pgt\Pgt$ signal and the $\PZ/\Pggx \to \Pq\Paq'$ background.
Since we expect the probability for a quark or gluon jet to be misidentified as $\tauh$ to be in the order to $10^{-2}$ or below,
it is particularly important to have sufficient background statistics.
The $\Pgt$ decays are simulated using \textsc{Pythia8}, and the generator tune and other settings are based on Ref.~\cite{Amhis:2021cfy}.
No $\Pgg\Pgg \to \mathrm{hadrons}$ overlay background is included in our simulation.
The study of this overlay is left to future work.

The stable particles from \textsc{Pythia8} are passed to a full \textsc{Geant4}~\cite{Agostinelli:2002hh}-based detector simulation and subsequent reconstruction based on the CLICdet detector~\cite{CLICdp:2017vju} and the \textsc{Marlin} reconstruction code~\cite{Gaede:2006pj}, interfaced in the \textsc{Key4HEP}~\cite{Ganis:2021vgv} software package.
We use the CLICdet detector (\verb+CLIC_o3_v14+), as the detector design has been thoroughly studied, and a rather complete implementation of tracker, calorimeter and particle flow reconstruction is available.
The CLICdet detector is optimized for precision physics and is based on a silicon pixel detector and tracker, a Si-W electromagnetic calorimeter and a scintillating hadronic calorimeter, encased in a 4T solenoid.
More details about the expected physics performance, including track and jet energy reconstruction properties, is available in~\cite{Abramowicz:2016zbo} and references therein.
Moreover, the CLICdet detector is conceptually similar to the proposed CLD detector for the FCC-ee~\cite{Bacchetta:2019fmz}, and thus a relevant benchmark model for particle identification and reconstruction studies.

Based on the \textsc{Marlin} reconstruction in \textsc{Key4HEP}, the output of the simulation and reconstruction chain is a set of \textsc{Pandora}~\cite{Marshall:2012hh,Marshall:2012ry,Marshall:2015rfa} particle flow candidates for each event, described by a four-momentum, a charge and a particle identification label: electron ($\Pe$), muon ($\Pgm$), photon ($\Pgg$), charged hadron ($\hpm$), and neutral hadron ($\hzero$).
The \textsc{Pandora} particle flow algorithm aggregates calorimeter hits to clusters and combines tracks and calorimeter clusters to reconstruct stable particle candidates.

The ParticleTransformer and DeepTau algorithms use the transverse ($\dxy$) and longitudinal ($\dz$) impact parameters of tracks to improve the discrimination of $\tauh$ and jets.
In jets arising from the hadronization of light quarks and gluons, the $\dxy$ and $\dz$ of the tracks are expected to be compatible with zero within their respective uncertainties ($\sigmaxy$ and $\sigmaz$),
while non-zero impact parameters are expected for the charged particles produced in $\Pgt$ decays,
reflecting the small distance that $\Pgt$ leptons travel between their production and decay.
The $\dxy$ and $\dz$ are not part of the \textsc{Key4HEP} format used in \textsc{Pandora} and thus need to be computed for the work presented in this paper.
As the distances that $\Pgt$ leptons travel between their production and decay are typically small compared to the expected radius of curvature of the tracks originating from the $\Pgt$ decay,
we simplify the task of computing the $\dxy$ and $\dz$ by using a linear approximation to the equations of motion for a charged particle in a magnetic field~\cite{Jackson:1999}.
Details of this approximation are documented in the repository of the code~\cite{christian_veelken_2023_8113344}.

The particle flow candidates are clustered to jets using the generalized $\kt$ algorithm for $\Pep\Pem$ collisions (\verb+ee_genkt+)~\cite{Boronat:2016tgd} with $p = -1$ and $R = 0.4$.
All jets considered in this paper are required to satisfy the conditions $10 < \theta < 170\degr$ and $\pT > 20$~\GeV,
where the symbol $\theta$ refers to the polar angle, with the $z$-axis taken to be the beam axis, and $\pT$ to the transverse momentum.
The condition on $\theta$ selects jets within the geometric acceptance of the tracking detector.
Due to the Lorentz boost in $\Pgt$ direction,
the particles produced in $\Pgt$ decays become more collimated in the detector as the energy of the $\Pgt$ lepton increases.
The selection $\pT > 20$~\GeV ensures that, in the signal samples, all particles produced in $\Pgt$ decays are within a narrow inner region of the jet.

The jets reconstructed in $\PZ/\Pggx \to \Pgt\Pgt$ and $\PZ\PH$, $\PH \to \Pgt\Pgt$ signal events are required to be matched to generator-level $\tauh$ within $\Delta R < 0.4$, while those reconstructed in $\PZ/\Pggx \to \Pq\Paq'$ background events are required to be matched to either a quark or a gluon on generator level.
The distance $\Delta R$ between generator-level and reconstructed particles is computed as:
\begin{equation}\label{eq:deltaR}
    \Delta R = \sqrt{(\theta_{i} - \theta_{j})^{2} + (\phi_{i} - \phi_{j})^{2}} \, ,
\end{equation}
where the symbol $\phi$ denotes the azimuthal angle of the jet, the subscript $i$ refers to the direction of the reconstructed jet, and the subscript $j$ to that of the 
generator-level $\tauh$, quark, or gluon.
Reconstructed jets that are close to generator-level electrons or muons are removed from the signal and background samples.
This leaves us with approximately $1.2$ and $0.9$ million reconstructed jets in the $\PZ/\Pggx \to \Pgt\Pgt$ and $\PZ\PH$, $\PH \to \Pgt\Pgt$ signal samples, respectively,
and $3.5$ million jets in the background sample.

For each jet, we store the associated particle-flow candidates as well as the jet four momentum.
The samples are shuffled and divided to mutually exclusive train-validation-test samples with a $26:9:65\%$ split.
The training dataset is used to optimize the parameters of the LorentzNet, ParticleTransformer, and DeepTau algorithms,
while the validation dataset is used to monitor the training.
The final performance of the different algorithms is evaluated using the test dataset.

\section{Tau identification algorithms}\label{sec:TauIDAlgos}

\begin{figure*}[ht!]
    \centering
    \ifx\ver\verPreprint
        \includegraphics[width=0.48\textwidth]{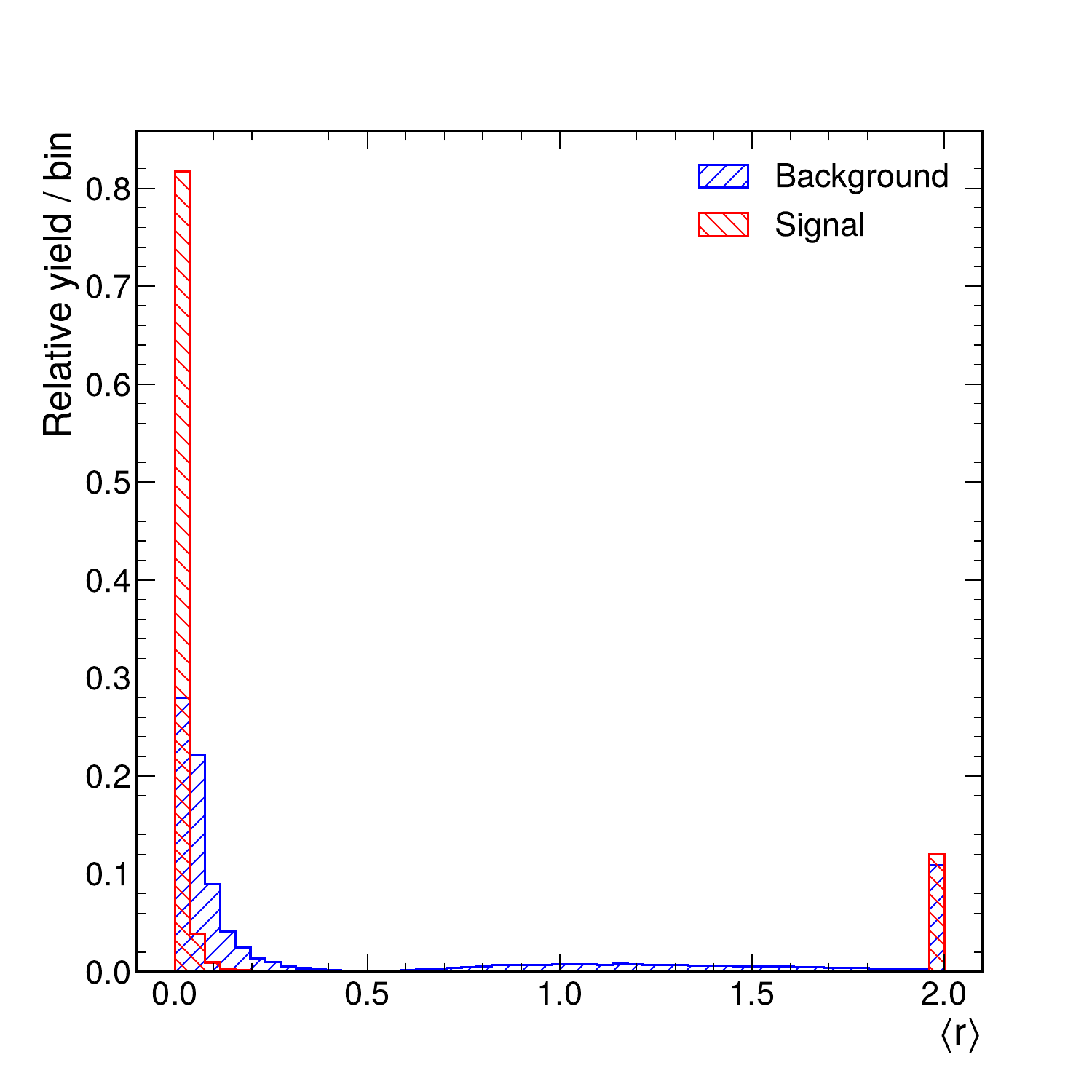}
        \includegraphics[width=0.48\textwidth]{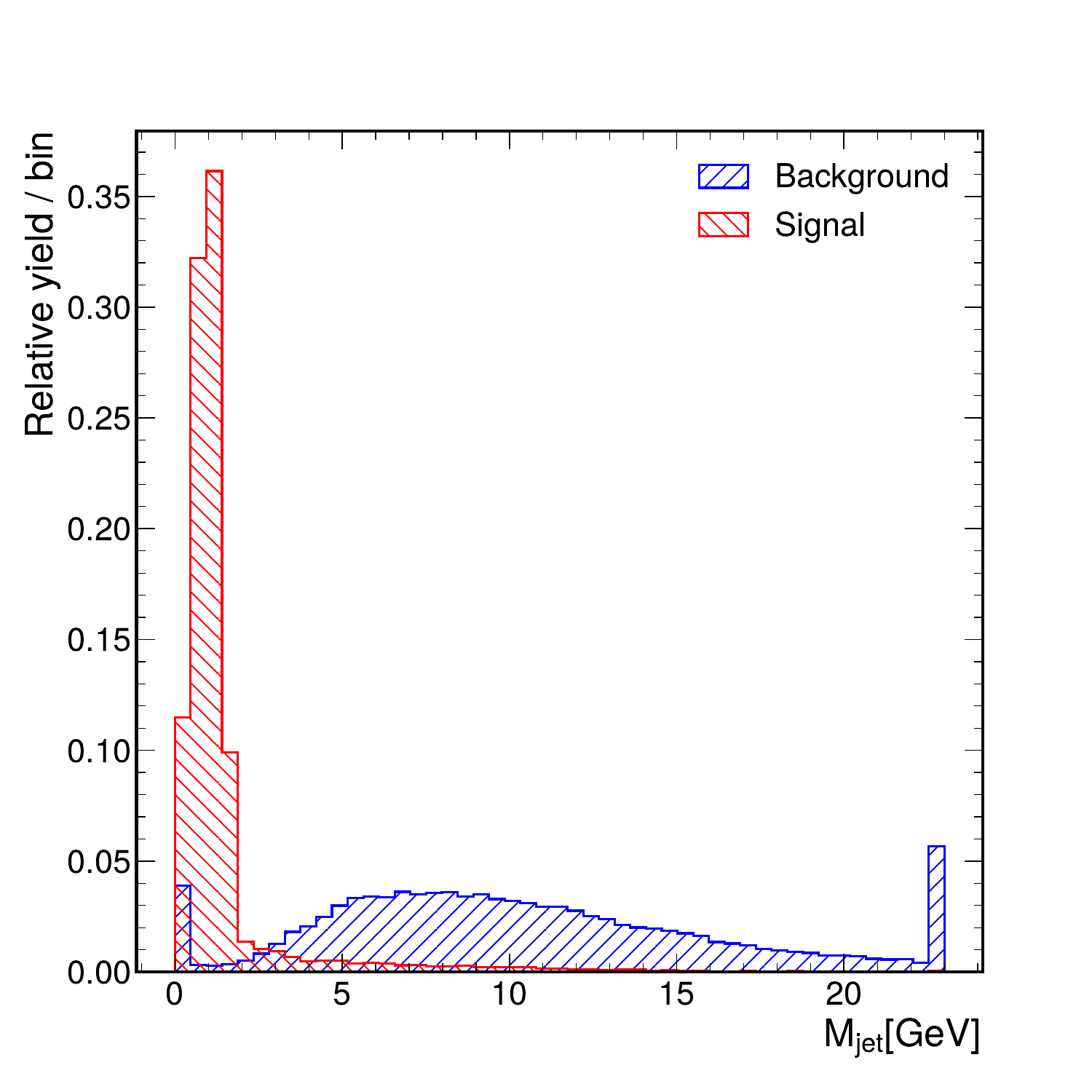}
        \includegraphics[width=0.48\textwidth]{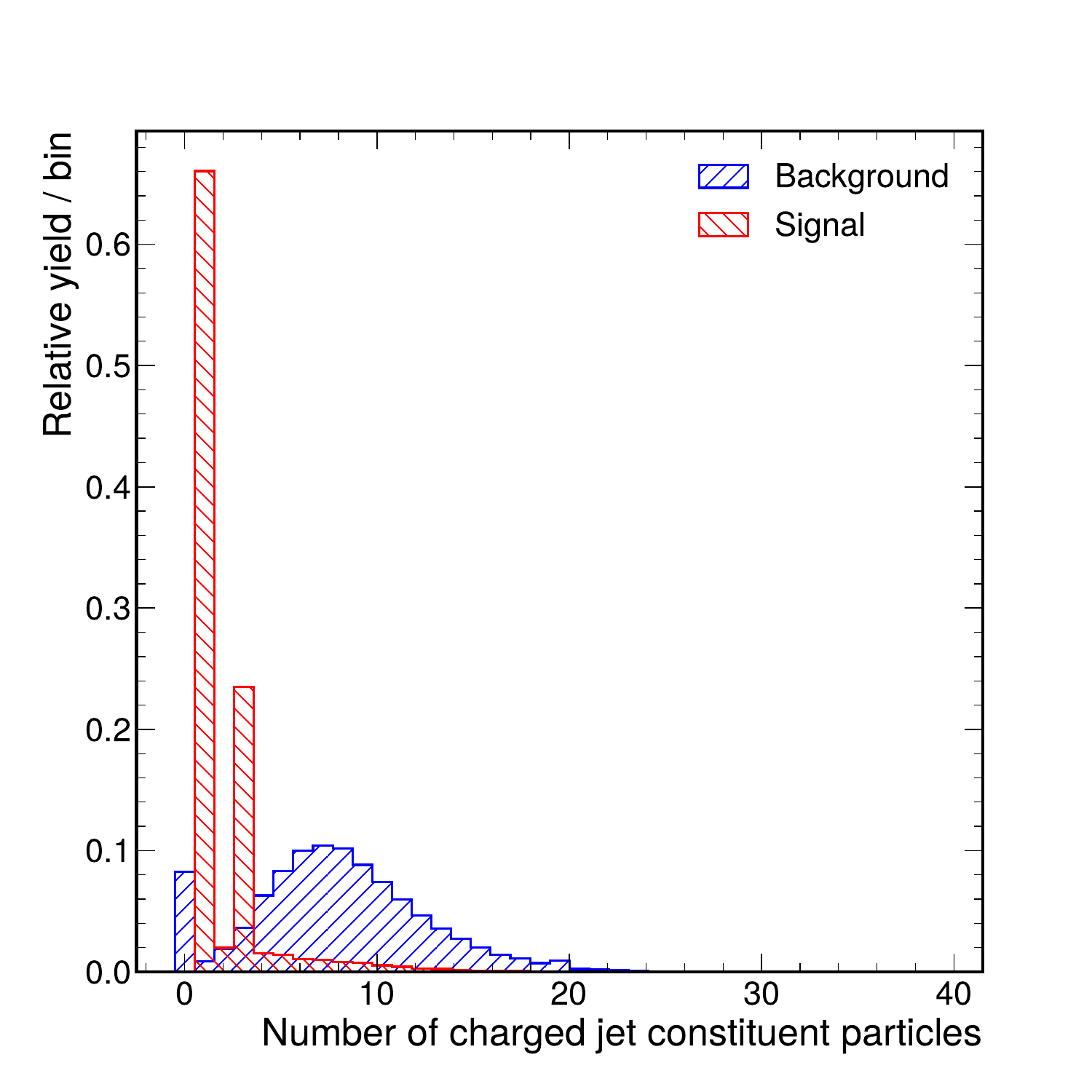}
        \includegraphics[width=0.48\textwidth]{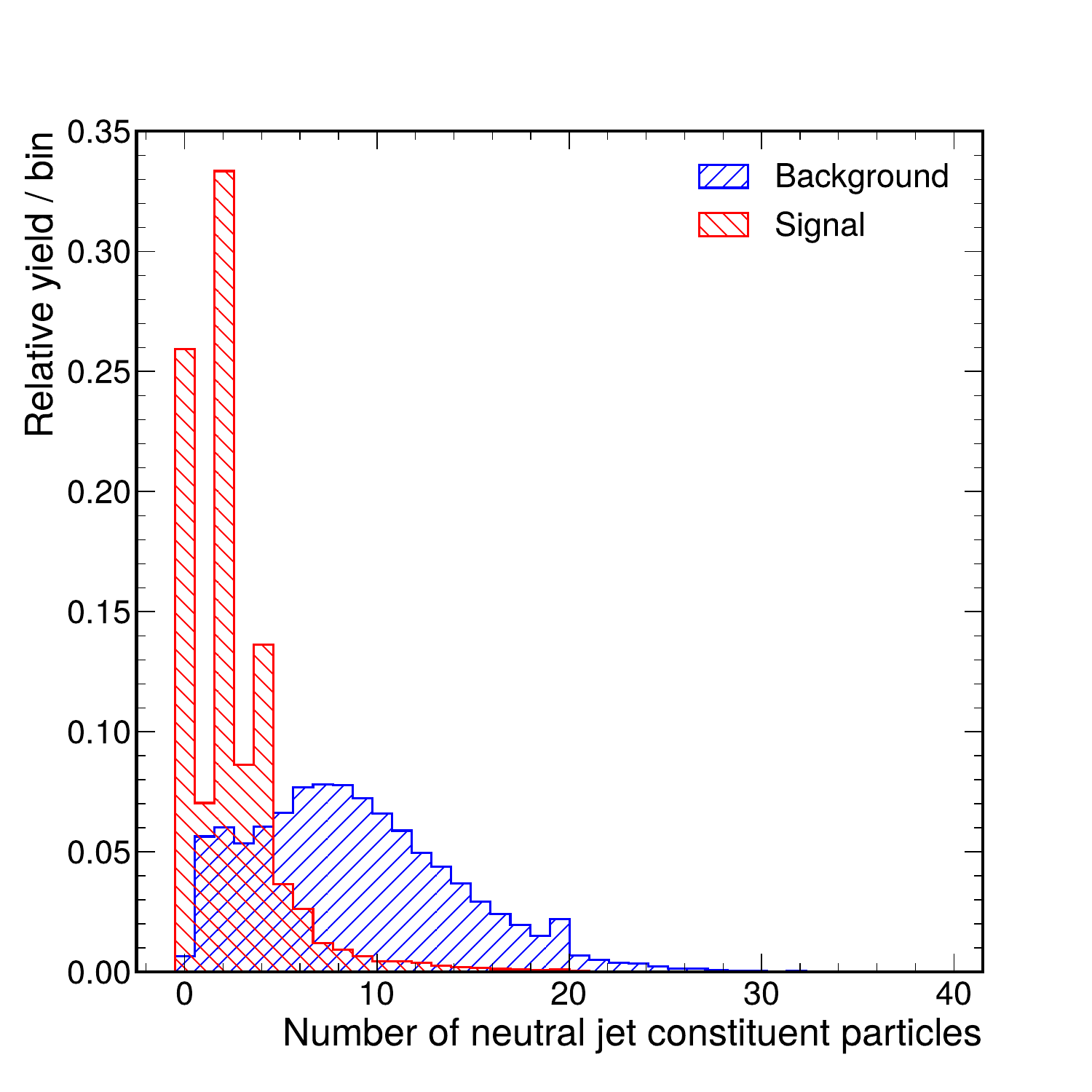}
    \fi
    \ifx\ver\verPAPER
        \includegraphics[width=0.43\textwidth]{Figs/QG_vs_tau/jet_radius.pdf}
        \includegraphics[width=0.43\textwidth]{Figs/QG_vs_tau/jet_mass.pdf}
        \includegraphics[width=0.43\textwidth]{Figs/QG_vs_tau/n_ch_particles_in_jet.pdf}
        \includegraphics[width=0.43\textwidth]{Figs/QG_vs_tau/n_neutral_particles_in_jet.pdf}
    \fi
    \caption{
      Distributions in the jet radius $\jetR$ (upper left), the mass $\jetM$ of the jet (upper right), and in the number $N$ of charged (lower left) and neutral (lower right) particles in the jet.
      The distributions are plotted on generator level for jets originating from hadronic $\Pgt$ decays (``signal'') compared to quark and gluon jets (``background''). 
      The rightmost bins of each distribution represent overflow bins.
    }
    \label{fig:discrObservables}
\end{figure*}

The discrimination between $\tauh$ from jets takes advantage of the fact that $\tauh$ are typically more collimated and contain fewer particles compared to quark and gluon jets. 
Distributions of the jet radius and of the number of particles in the jet are shown in Fig.~\ref{fig:discrObservables} for illustration.
The jet radius $\jetR$ is defined by:
\begin{equation}\label{eq:jetRadius}
    \jetR = \frac{\mathlarger{\sum}_{i} \, \pT^{i} \, \, \Delta R}{\mathlarger{\sum}_{i} \, \pT^{i}},
\end{equation}
where the sum extends over all particles $i$ within the jet and the distance $\Delta R$ between the direction of the jet $j$ and particle $i$ is given by Eq.~(\ref{eq:deltaR}).
We also show the distribution in jet mass. Quark and gluon jets typically have a higher mass than $\tauh$,
reflecting the higher particle multiplicity and wider angular spread,
while the mass of $\tauh$ is bounded from above by the $\Pgt$ lepton mass of $1.777$~\GeV~\cite{ParticleDataGroup:2022pth}.

Each $\tauh$ identification algorithm considered in this paper
is seeded by the collection of jets described in Section~\ref{sec:MCSamples_and_event_reconstruction} and uses the constituent particles of these jets as input to the $\tauh$ reconstruction.
The output of each algorithm is a discriminant $\Dtau$ in the range $0$ to $1$,
where a value of $1$ means that the jet is identified as $\tauh$, while a value of $0$ means that the jet is identified as originating from the hadronization of either a quark or a gluon.

All constituents of the jet are considered in the $\tauh$ reconstruction.
No selection on $\theta$ and $\pT$ of the jet constituents is applied, because we observed that not applying such selection improves (by a small amount) the $\tauh$ identification performance.
We remark that it may be necessary to apply a $\pT$ threshold on the jet constituents
in order to reduce the effect of the $\gamma \gamma \to \mathrm{hadrons}$ overlay background, which is not included in our simulation, 
or in case particles of low $\pT$ are not well modeled by the MC simulation.

The performance of each algorithm is evaluated in terms of $\tauh$ identification efficiency and of the misidentification rate for quark and gluon jets.
We denote the former by the symbol $\efficiency$ and the latter by $\fakerate$.
The $\tauh$ identification efficiency corresponds to the probability for a genuine $\tauh$ in the $\PZ/\Pggx \to \Pgt\Pgt$ and $\PZ\PH$, $\PH \to \Pgt\Pgt$ signal samples
to pass a selection on the discriminant $\Dtau$, while the misidentification rate refers to the probability for jets 
that originate from the hadronization of a quark or gluon in the $\PZ/\Pggx \to \Pq\Paq'$ background sample to pass this selection.
The probability is defined by:
\begin{equation}\label{eq:efficiency_and_fakerate}
    \mathcal{P} = \frac{\pT^{\rec} > 20\mbox{~\GeV} \conjunction 10 < \theta_{\rec} < 170\degr \conjunction \mathcal{D}_{\Pgt} > \mathcal{T}}{\pT^{\genX} > 20\mbox{~\GeV} \conjunction 10 < \theta_{\genX} < 170\degr} \, ,
\end{equation}
where the symbols $\mathcal{P}$ and $\X$ refer to the efficiency $\efficiency$ (to the misidentification rate $\fakerate$)
and to the system of charged hadrons and neutral pions produced in the $\Pgt$ decay (to the quark or gluon that initiated the jet) in case of signal (background).
The symbols $\pT^{\rec}$ and $\theta_{\rec}$ refer to the transverse momentum and polar angle of the jet that seeds the $\tauh$ reconstruction in case of the LorentzNet and ParticleTransformer algorithms and to the $\pT$ and $\theta$ of the $\tauh$ object reconstructed by the HPS algorithm in case of the DeepTau algorithm.
The symbol ``$\&$'' denotes conjunction and $\mathcal{T}$ refers to the threshold imposed on the discriminant $\mathcal{D}_{\Pgt}$.
The acceptance criteria defined by the denominator are applied in the numerator also.
The efficiency $\efficiency$ as well as the misidentification rate $\fakerate$ depend on the threshold $\mathcal{T}$ and in general vary with $\pT$ and $\theta$.

\subsection{LorentzNet}

The LorentzNet algorithm~\cite{Gong:2022lye} employs a DNN architecture based on the attention mechanism~\cite{bahdanau2016neural,vaswani2017attention}.
The algorithm uses as input the four-momentum, type, and charge of the $N$ particles of highest $\pT$ among the jet's particle constituents, plus the two beam particles (the colliding $\Pep$ and $\Pem$).
In case the jet contains fewer than $N$ particles, the ``missing'' particles are represented by zeros.
The number $N$ of input particles constitutes a parameter of the algorithm, which needs to be chosen by the user. 
Based on Fig.~\ref{fig:discrObservables}, we choose $N = 25$.
The particle type ($\Pe$, $\Pgm$, $\Pgg$, $\hpm$, $\hzero$) is passed in one-hot encoded format~\cite{scikit-learn}. The architecture of the DNN is designed such that the output of the algorithm is equivariant to proper orthochronous Lorentz transformations (translations, rotations, and boosts) of the particles' four-momenta and is invariant to permutations of any pair of particles. The latter property means that the LorentzNet algorithm is invariant to the ordering in which the jet's particle constituents are presented to the algorithm. A function $\phi: \R^{n} \rightarrow \R^{m}$ is equivariant to actions of the Lorentz group $G$ if the following relationship holds: $\phi(g(x)) = g(\phi(x))$, where $x \in \R^{n}$, $\phi(x) \in \R^{m}$, and $g \in G$.
The Lorentz equivariance introduces an inductive bias into the algorithm, with the aim of improving the algorithm's capability for generalization.
It is demonstrated in Refs.~\cite{villar2023scalars,Gong:2022lye} that this DNN architecture is flexible enough to approximate any Lorentz equivariant function. 
The algorithm is implemented in \textsc{PyTorch}~\cite{paszke2019pytorch}. All parameters of the LorentzNet algorithm, which need to be chosen by the user, are set to the values given in Section 3.3 of Ref.~\cite{Gong:2022lye}, except for the parameter $N$, as explained above, and the parameter $c$, which we set to $c = 0.005$.

The training is performed in batches of $128$ jets and for a maximum of $100$ epochs. The algorithm has $2.3 \cdot 10^{5}$ trainable parameters, which are updated after each epoch using the AdamW~\cite{loshchilov2019decoupled} optimizer, in order to minimize the loss on the training dataset.
The learning rate is varied according to the one-cycle policy~\cite{smith2018superconvergence} during the training, with the maximum learning rate set to $10^{-3}$.
The focal loss from Ref.~\cite{Lin:2017fqe} with $\gamma = 2$ is used for the loss function. We found that this choice of loss function improves the separation of $\tauh$ from quark and gluon jets in particular for intermediate values of the DNN output, compared to using binary cross-entropy loss.
The loss on the validation dataset is computed after each training epoch. The final DNN parameters are chosen to be those that minimize the loss on the validation dataset.

\begin{figure}[ht!]
    \centering
    \ifx\ver\verPreprint
        \includegraphics[width=0.52\textwidth]{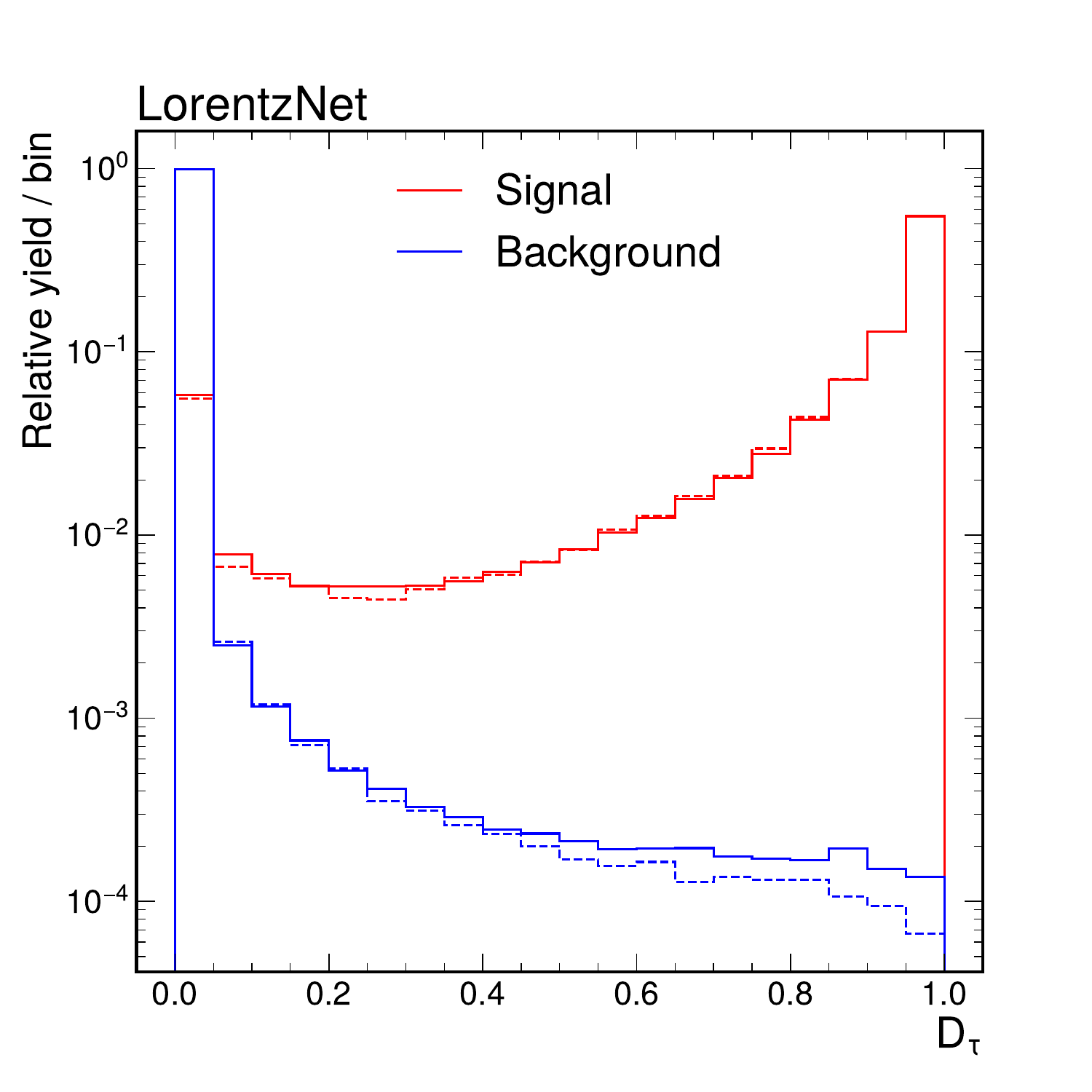}
    \fi
    \ifx\ver\verPAPER
        \includegraphics[width=0.95\columnwidth]{Figs/tauClassifiers/LorentzNet_tauClassifier.pdf}
    \fi
    \caption{
      Distribution in the discriminant $\Dtau$ for the LorentzNet algorithm.
      The solid curves refer to the test dataset and the dashed curves to the training dataset.  
    }
    \label{fig:tauClassifier_LorentzNet}
\end{figure}

Distributions in the discriminant $\Dtau$ for the training and test datasets are shown in Fig.~\ref{fig:tauClassifier_LorentzNet}. The distributions on the test sample are represented by solid lines, while the dashed lines represent the distributions on the training sample. A moderate amount of overtraining can be seen in the figure. The main effect of the overtraining is that the tail of the background distribution in the region of high values of the discriminant $\Dtau$ is more pronounced for the test sample than for the training sample.

\subsection{Particle Transformer}

The ParticleTransformer algorithm~\cite{Qu:2022mxj} is also based on the attention mechanism~\cite{bahdanau2016neural,vaswani2017attention}. Its architecture has originally been developed in the context of natural-language processing and is referred to as Transformer model in the literature~\cite{devlin2019bert,GPT}. 
The ParticleTransformer extends the LorentzNet algorithm by using additional input variables, notably the transverse and longitudinal impact parameters of charged particles, plus four observables, which represent properties of particle pairs. The algorithm is implemented in \textsc{PyTorch}~\cite{paszke2019pytorch}.
For the per-particle features, we use the $17$ observables given in Table~2 of Ref.~\cite{Qu:2022mxj}. The four pairwise features are the distance $\Delta R$ between the particles, given by Eq.~(\ref{eq:deltaR}), the mass of the particle pair, and the two observables:
\ifx\ver\verPreprint
\begin{equation}
\kt = \min\left( \pT^{i}, \pT^{j} \right) \, \Delta R \quad \text{and} \quad z = \min\left( \pT^{i}, \pT^{j} \right) / \left( \pT^{i} + \pT^{j} \right) \, ,
\end{equation}
\fi
\ifx\ver\verPAPER
\begin{eqnarray}
    \kt & = & \min\left( \pT^{i}, \pT^{j} \right) \, \Delta R \nonumber \\
    z & = & \min\left( \pT^{i}, \pT^{j} \right) / \left( \pT^{i} + \pT^{j} \right) \, ,
\end{eqnarray}
\fi
where the superscripts $i$ and $j$ refer to the first and second particle of the pair, respectively.
The $25$ particles of highest $\pT$ among the particle constituents of the jet are considered when computing the per-particle and pairwise features.

The algorithm is trained in batches of $128$ jets for a maximum of $100$ epochs.
The $2.1 \cdot 10^{6}$ trainable parameters are updated after each epoch, using the AdamW~\cite{loshchilov2019decoupled} optimizer.
The other training parameters are the same as for the LorentzNet algorithm.
The final DNN parameters are taken to be those that minimize the loss on the validation dataset, which is computed after each epoch.

\begin{figure}[ht!]
    \centering
    \ifx\ver\verPreprint
        \includegraphics[width=0.52\textwidth]{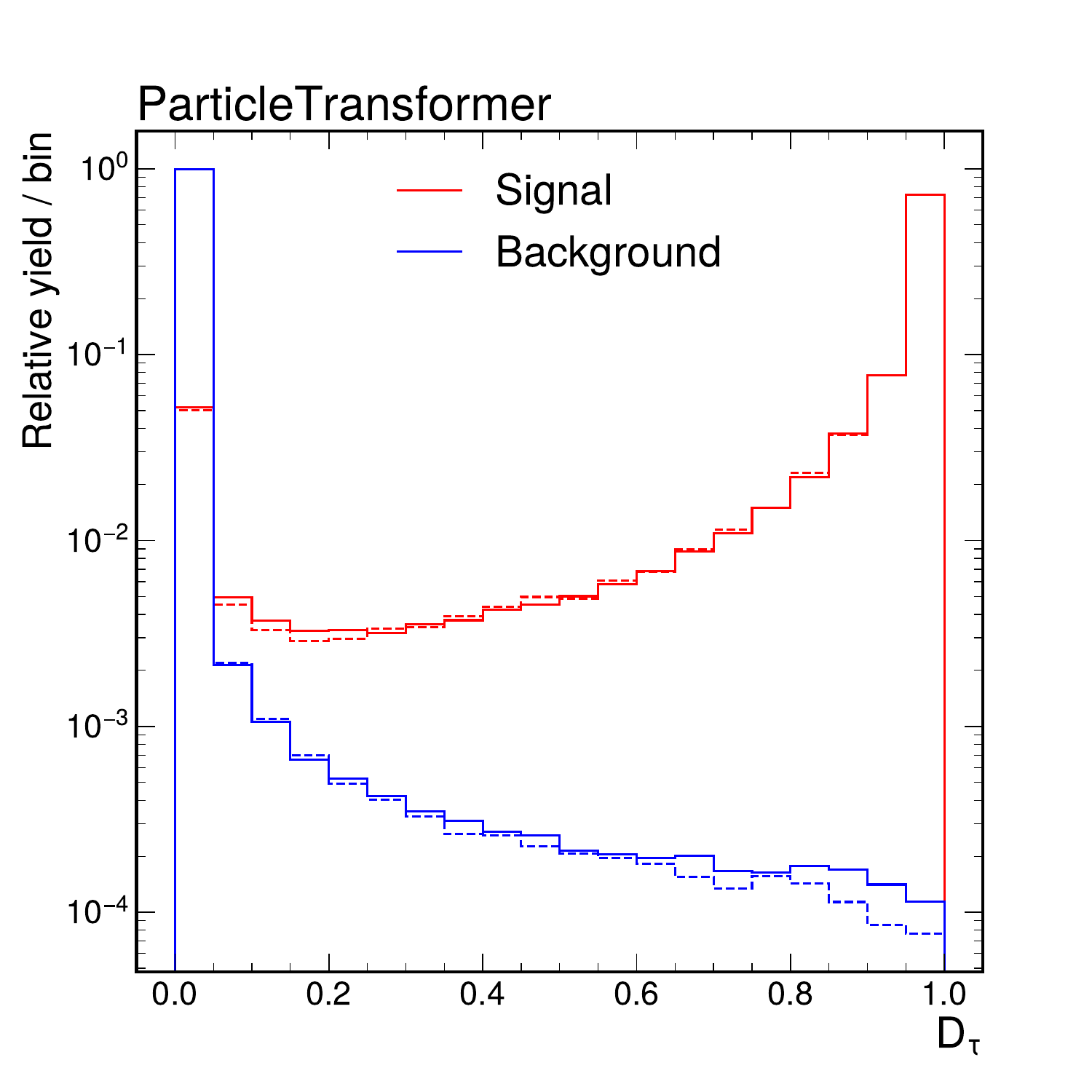}
    \fi
    \ifx\ver\verPAPER
        \includegraphics[width=0.95\columnwidth]{Figs/tauClassifiers/ParticleTransformer_tauClassifier.pdf}
    \fi
    \caption{
      Distribution in the discriminant $\Dtau$ for the ParticleTransformer algorithm.
      The solid curves refer to the test dataset and the dashed curves to the training dataset.  
    }
    \label{fig:tauClassifier_ParticleTransformer}
\end{figure}

Distributions in the discriminant $\Dtau$ computed by the ParticleTransformer algorithm are shown in Fig.~\ref{fig:tauClassifier_ParticleTransformer}.
A moderate amount of overtraining, similar in magnitude and shape effect to that of the LorentzNet algorithm, can be seen in the figure.

\section{Results}
\label{sec:Results}

\begin{figure}[ht!]
    \centering
    \ifx\ver\verPreprint
        \includegraphics[width=0.56\textwidth]{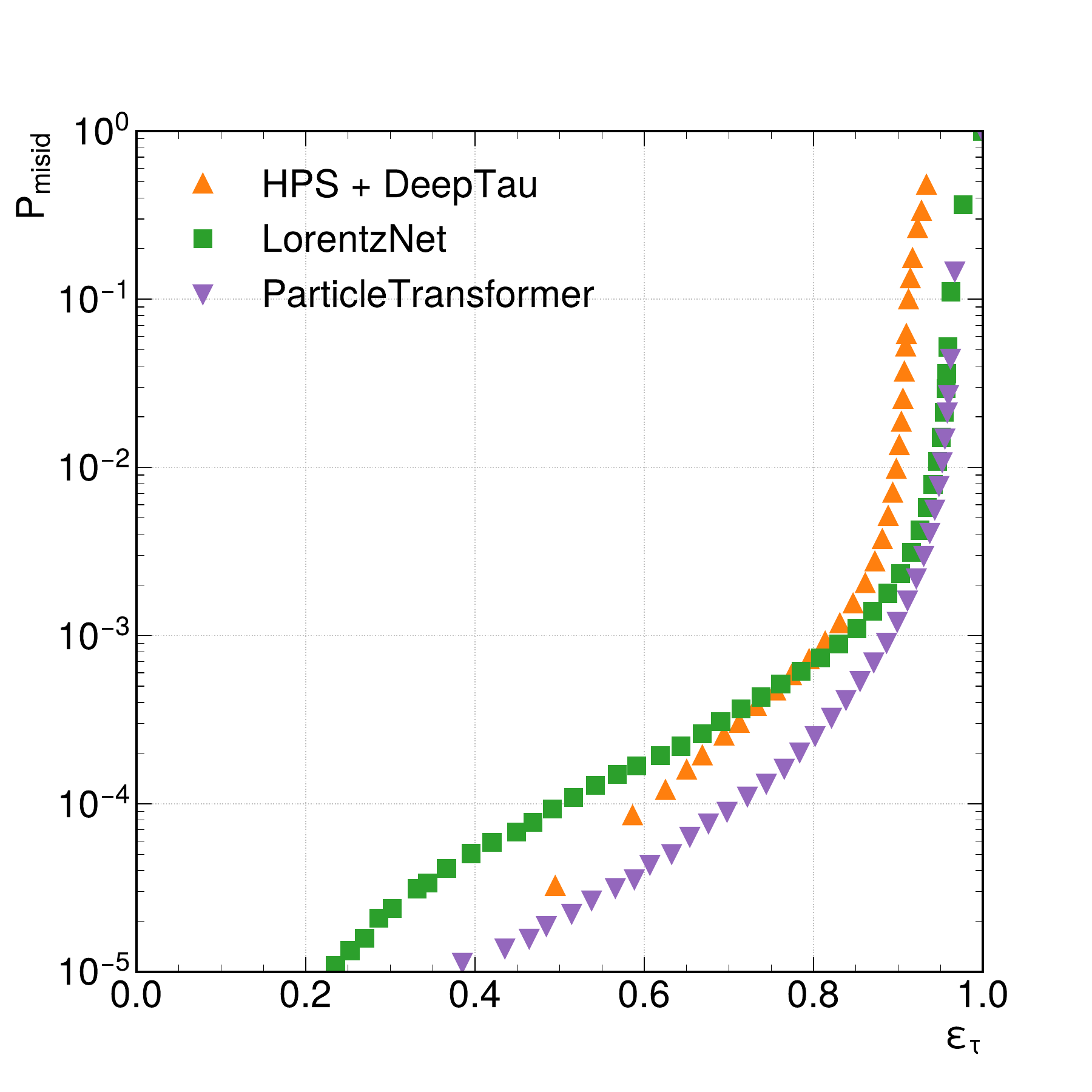}
    \fi
    \ifx\ver\verPAPER
        \includegraphics[width=0.98\columnwidth]{Figs/performance/ROC.pdf}
    \fi
    \caption{
      Misidentification rate for quark and gluon jets as function of the $\tauh$ identification efficiency for the LorentzNet and ParticleTransformer algorithms,
      compared to the combination of the HPS + DeepTau algorithm.
    }
    \label{fig:rocCurves}
\end{figure}

The ``receiver operating characteristic'' (ROC) curves of the LorentzNet and ParticleTransformer algorithms is compared to the one of the DeepTau algorithm in Fig.~\ref{fig:rocCurves}.
The ROC curves show the $\tauh$ identification efficiencies $\efficiency$ for the $\PZ\PH$, $\PH \to \Pgt\Pgt$ signal sample on the axis of abscissas
and the misidentification rates $\fakerate$ for the $\PZ/\Pggx \to \Pq\Paq'$ background sample on the ordinate.
The curves are constructed by varying the threshold imposed on $\Dtau$ in $1000$ steps within the range $0$ to $1$ 
and computing $\efficiency$ and $\fakerate$ according to Eq.~(\ref{eq:efficiency_and_fakerate}) for each such threshold.
Points on the left side of the curve correspond to a tighter selection on the output $\Dtau$ of the $\tauh$ identification algorithm,
while points on the right side correspond to a looser selection.
All three algorithms achieve misidentification rates on the level of a permille or below for $\tauh$ identification efficiencies in the range $50$-$80\%$, 
the range we expect to be most relevant for physics analyses.
Numerical values for $\fakerate$ for $\efficiency$ values of $50$, $60$, $70$, and $80\%$ are given in Table~\ref{tab:performance}.

\begin{table*}[ht!]
    \centering
    \begin{tabular}{lr@{$ \,\,\pm\,\, $}rr@{$ \,\,\pm\,\, $}rr@{$ \,\,\pm\,\, $}rr@{$ \,\,\pm\,\, $}rr@{$ \,\,\pm\,\, $}r}
        Algorithm & \multicolumn{2}{c}{$\langle \varepsilon_{\Pgt} \rangle = 50\%$} & \multicolumn{2}{c}{$\langle \varepsilon_{\Pgt} \rangle = 60\%$} & \multicolumn{2}{c}{$\langle \varepsilon_{\Pgt} \rangle = 70\%$} & \multicolumn{2}{c}{$\langle \varepsilon_{\Pgt} \rangle = 80\%$} \\
        \hline
        HPS + DeepTau       & $31034$ & $3849$ & $11796$ & $902$ & $3564$ & $150$ & $1308$ & $33$ \\
        LorentzNet          & $10137$ & $719$ & $5634$ & $298$ & $3006$ & $116$ & $1450$ & $39$ \\
        \textbf{ParticleTransformer} & $\mathbf{48028}$ & $\mathbf{7411}$ & $\mathbf{24 904}$ & $\mathbf{2 767}$ & $\mathbf{10617}$ & $\mathbf{770}$ & $\mathbf{4 042}$ & $\mathbf{181}$ \\
    \end{tabular}
    \caption{
      Average misidentification rates, computed according to Eq.~(\ref{eq:efficiency_and_fakerate}),
      for average $\tauh$ identification efficiencies $\langle \varepsilon_{\Pgt} \rangle$ of $50$, $60$, $70$, and $80\%$.
      The numbers given in the table correspond to the inverse, $1/\langle P_{\misid} \rangle$, of the average misidentification rates, and to the statistical uncertainties on these values. 
      Bold numbers highlight the best-performing algorithm. 
    }
    \label{tab:performance}
\end{table*}

The $\efficiency$ obtained for the $\PZ/\Pggx \to \Pgt\Pgt$ signal sample agrees with the one obtained for the $\PZ\PH$, $\PH \to \Pgt\Pgt$ sample
within $5$-$10\%$, depending on the threshold on $\Dtau$, where the quoted values refer to relative differences.
The differences between the signal samples are rather small, indicating that the reweighting described in Section~\ref{sec:MCSamples_and_event_reconstruction} works as intended
and the algorithms do not exploit differences in event kinematics.

The ParticleTransformer algorithm is seen to outperform the LorentzNet algorithm as well as the DeepTau algorithm,
achieving $\fakerate$ of $2.1 \times 10^{-5}$ and $2.5 \times 10^{-4}$ for $\efficiency$ of $50$ and $80\%$, respectively.
This performance demonstrates the potential of applying the advanced DL algorithms originally developed for jet-flavour tagging
to the task of $\tauh$ identification.

\begin{figure}[ht!]
    \centering
    \ifx\ver\verPreprint
        \includegraphics[width=0.48\textwidth]{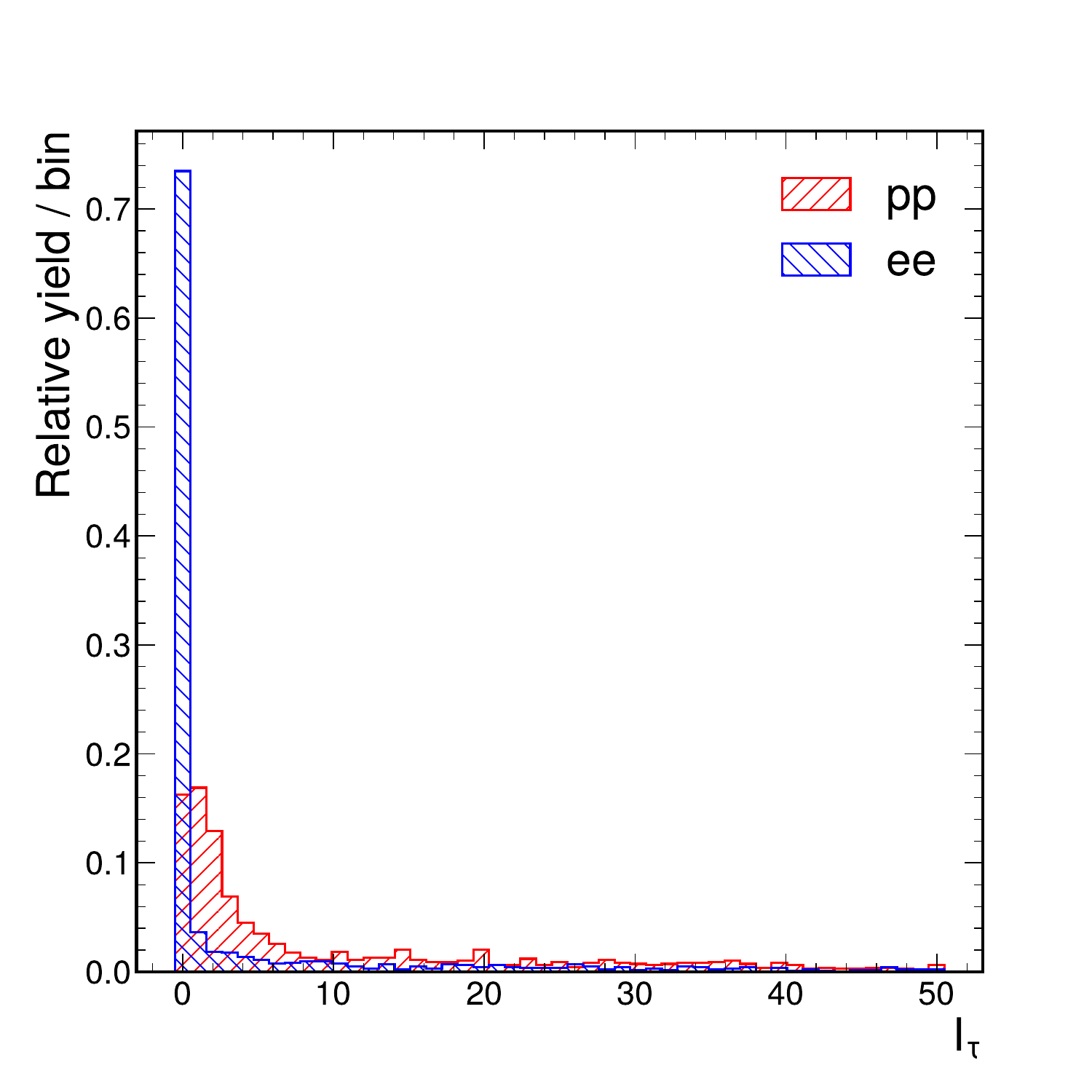}
        \includegraphics[width=0.48\textwidth]{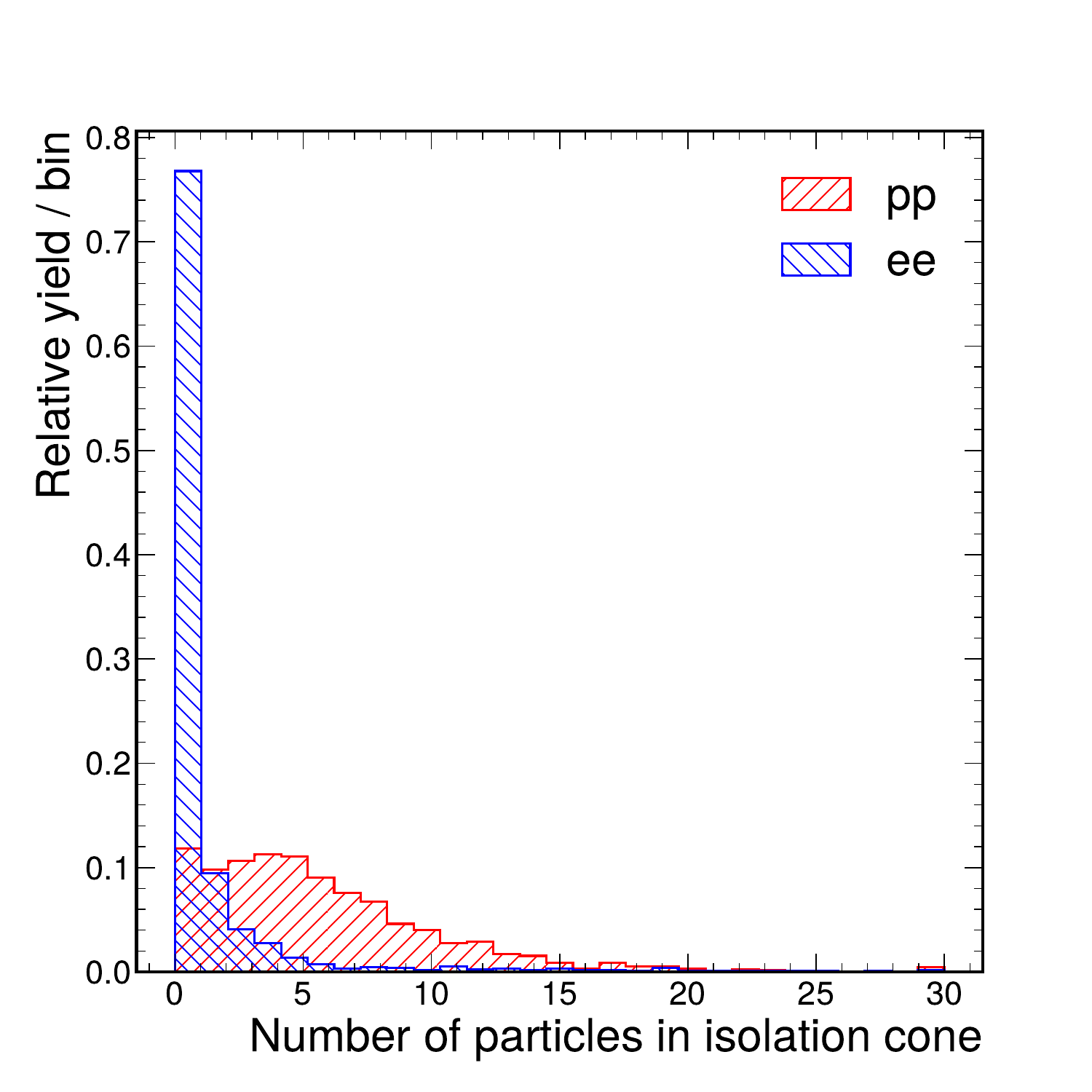}
    \fi
    \ifx\ver\verPAPER
        \includegraphics[width=0.90\columnwidth]{Figs/isolations.pdf}
        \includegraphics[width=0.90\columnwidth]{Figs/n_constituents.pdf}
    \fi
    \caption{
      Distributions in the isolation $I_{\Pgt}$ of the $\tauh$ and in the multiplicity $N_{\iso}$ of particles in the isolation cone of the $\tauh$, for hadronic $\Pgt$ decays in $\PZ\PH$, $\PH \to \Pgt\Pgt$ signal events produced in $\Pp\Pp$ (red) and in $\Pep\Pem$ (blue) collisions.
    }
    \label{fig:experimental_environment}
\end{figure}

We remark that the performance of the HPS + DeepTau algorithm in our study is significantly higher than the performance reported in Ref.~\cite{CMS:2022prd}.
The higher performance is reflected by the $\fakerate$ values for the DeepTau algorithm shown in Fig.~\ref{fig:rocCurves},
which are about an order of magnitude lower compared to the misidentification rates shown in Fig.~4 of Ref.~\cite{CMS:2022prd},
for similar $\tauh$ identification efficiencies.
We believe the main reason for this difference to be the ``cleaner'' experimental environment of $\Pep\Pem$ compared to $\Pp\Pp$ collisions,
which considerably simplifies the task of discriminating $\tauh$ from quark and gluon jets.
The ROC curve shown in Fig.~4 of Ref.~\cite{CMS:2022prd} was made for typical experimental conditions during LHC Run $2$,
which, besides the general higher hadronic activity arising from the production of extra jets and the underlying event,
also included about $50$ minimum bias overlay interactions, referred to as ``pileup''~\cite{LUM-21-001}.
To illustrate this point, we show in Fig.~\ref{fig:experimental_environment} the distributions in the number of particles $N_{\iso}$ in the isolation cone 
and in the observable $I_{\Pgt} = I_{\textrm{ch5}} + I_{\Pgg5}$, where $I_{\textrm{ch5}}$ and $I_{\Pgg5}$ are computed according to Eq.~(\ref{eq:Itau}) in the appendix,
for $\tauh$ in $\PZ\PH$, $\PH \to \Pgt\Pgt$ events produced in $\Pp\Pp$ collisions at $\sqrt{s} = 13$~\TeV with $50$ pileup interactions
and in $\PZ\PH$, $\PH \to \Pgt\Pgt$ events produced in $\Pep\Pem$ collisions at $\sqrt{s} = 380$~\GeV.
Particles that are matched to $\Pgt$ decay products on generator-level are excluded from the computation of the observables $N_{\iso}$ and $I_{\Pgt}$.
The observable $I_{\Pgt}$ represents one of the main handles to separate $\tauh$ from quark and gluon jets and significantly benefits from the ``cleaner'' experimental environment.

\section{Summary and outlook}
\label{sec:Summary}

The main result of this paper is that the advancements in modern deep-learning techniques, which drove the recent progress in jet-flavour tagging,
can similarly be applied to the task of identifying hadronic $\Pgt$ decays.
Of the two jet-flavour tagging algorithms that we studied, LorentzNet and ParticleTransformer,
the ParticleTransformer algorithm provides the superior performance.
It achieves a misidentification rate of $2.1 \times 10^{-5}$ ($2.5 \times 10^{-4}$) for a $\tauh$ identification efficiency of $50$ ($80\%$).
A low misidentification rate is of particular importance for measurements of $\Pgt$ lepton branching fractions via the ``tag-and-probe'' method,
as discussed in Section 1 of Ref.~\cite{Pich:2020qna}.
Remarkably, the ParticleTransformer algorithm achieves this level of performance while performing $\tauh$ reconstruction and identification in a single step without any manual tuning by domain experts.
We believe this result, that algorithms originally developed for jet-flavour tagging and repurposed for the task of $\tauh$ identification
provide a performance that is as good as or better than state-of-the-art $\tauh$ identification algorithms to be applicable to $\Pp\Pp$ collisions at the LHC also.
We remark that the performance of the DL-based algorithms is achieved in an end-to-end approach without manual tuning by domain experts,
while in particular the HPS algorithm benefitted from substantial tuning by domain experts.

We believe the numerical values for $\efficiency$ and $\fakerate$ given in Table~\ref{tab:performance} may be useful in the context
of sensitivity studies of physics analyses with $\Pgt$ leptons at future high-energy $\Pep\Pem$ experiments,
similar to how the parametrizations of particle reconstruction and identification performances provided in the \textsc{Delphes}~\cite{deFavereau:2013fsa} fast detector simulation software
by the ATLAS and CMS experiments for $\Pp\Pp$ collisions at the LHC have been used.
We remark that $\efficiency$ and $\fakerate$ depend on $\pT$ and $\theta$.
This dependency is not reflected in the numbers given in Table~\ref{tab:performance},
which represent averages over the $\pT$ and $\theta$ spectrum of $\tauh$ and jets in the signal and background samples.
Values of $\efficiency$ and $\fakerate$ as function of $\pT$ and $\theta$ can be obtained from the authors upon request.

Owing to the ``cleaner'' experimental environment, the misidentification rates are substantially lower in $\Pep\Pem$ collisions compared to $\Pp\Pp$ collisions at the LHC, 
for similar $\tauh$ identification efficiencies.

Future work includes to study in detail the effect of the $\Pgg\Pgg \to \mathrm{hadrons}$ overlay background on the $\tauh$ identification performance
and to extend the ParticleTransformer algorithm to reconstruct individual hadronic $\Pgt$ decay modes.
The capability to distinguish individual hadronic $\Pgt$ decay modes is important for measurements of the $\Pgt$ spin polarization~\cite{Behnke:2013lya,Tran:2015nxa,Xu:2017lgs,Dam:2021ibi,Atag:2010ja}.
Furthermore, in analogy with jet-flavour tagging, it may be important to study the robustness of the taggers to theoretical and experimental systematic effects
and to design ML models that are robust against such effects.
Another important aspect for future work is hardware portability, latency and throughput, 
to ensure that ML-based reconstruction models are resource-efficient and can be deployed at the trigger level, if needed.

\section*{Acknowledgments}
The authors would like to thank Patrizia Azzi for suggesting to study the topic of $\tauh$ identification at future $\Pep\Pem$ colliders
and the authors of Ref.~\cite{CMS:2022prd} for help with the preparation of Fig.~\ref{fig:DeepTau_architecture}.
This work was supported by the Estonian Research Council grants PRG445 and PSG864.
Computing resources were provided by the NICPB HEP cluster.

\section*{Data availability}
The code used to produce the samples and perform the analysis is available at Ref.~\cite{christian_veelken_2023_8113344}.

\section*{Authorship statement}
TL implemented the reconstruction of transverse and longitudinal impact parameters.
SN implemented and trained the DeepTau algorithm. 
JP produced raw simulation samples and performed supervision and funding acquisition. 
LT developed and ran the software infrastructure to process raw simulation samples, performed the validation of the samples, and produced tables and figures for the paper. 
CV implemented the cut-based HPS algorithm, implemented and trained the LorentzNet and ParticleTransformer algorithms, performed supervision and funding acquisition, and took the lead role in preparation of the manuscript. 
All authors contributed text to the manuscript and provided significant input to the discussion.

\appendix
\section*{Appendix}

\subsection*{HPS algorithm}

The HPS algorithm is published in Refs.~\cite{CMS:2015pac,CMS:2018jrd}.
The algorithm aims to reconstruct individual hadronic $\Pgt$ decay modes.
The decay modes targeted by the HPS algorithm are $\Pgtm \to \hminus\Pgngt$, 
$\Pgtm \to \hminus\Pgpz\Pgngt$, $\Pgtm \to \hminus\Pgpz\Pgpz\Pgngt$,
$\Pgtm \to \hminus\hplus\hminus\Pgngt$, $\Pgtm \to \hminus\hplus\hminus\Pgpz\Pgngt$,
and the charge conjugates of these decays for $\Pgt^{+}$.

The $\Pgpz$ mesons are reconstructed by clustering photons of $\pT > 1.0$~\GeV into ``strips''. Electrons and positrons of $\pT > 0.5$~\GeV are included in the clustering, assuming that they originate from photon conversions within the tracking detector. The clustering proceeds via an iterative procedure. The procedure is seeded by the $\Pgg$ or $\Pe$ of highest $\pT$ that is not yet included in a strip,
where we use the symbol $\Pe$ to refer to $\Pem$ and $\Pep$ irrespective of their electric charge. The $\theta$ and $\phi$ of the seed defines the initial location of the strip. The $\Pgg$ or $\Pe$ of next highest $\pT$, which is within an $\theta \times \phi$ window of size $0.05 \times 0.20$ around the strip location, is added to the strip. The strip momentum is recomputed as the momentum sum of all particles in the strip and the position of the window is updated accordingly. The energy of the strip is chosen such that its mass matches the $\Pgpz$ meson mass. The clustering continues until there are no more unclustered $\Pgg$ or $\Pe$ within the $\eta \times \phi$ window. The algorithm then proceeds by choosing a new seed and building the next cluster. The reconstruction of $\Pgpz$ mesons ends when all jet constituents of type $\Pgg$ or $\Pe$ are clustered into strips.

The strips are combined with jet constituents of types $\hpm$ in the next step of the HPS algorithm. Following the implementation of the HPS algorithm in CMS, particles of type electron are also considered as ``charged hadrons'' in this step. Potential double-counting of reconstructed electrons is resolved at a later stage of the algorithm. The motivation for considering reconstructed $\Pe$ as $\hpm$ is that these $\Pe$ may result from the overlap of a charged hadron of low energy with a high energetic $\Pgpz$ meson. The experimental signature of such an overlap is a track that spatially matches a cluster in the electromagnetic calorimeter (ECAL) whose energy is significantly higher than the track momentum and little energy in the hadronic calorimeter (HCAL). This case often gets reconstructed as one particle of type electron by the PF algorithm in CMS. 

The combination of $\hpm$ with strips proceeds via a combinatorial approach.
A set of $\tauh$ candidates corresponding to combinations of either one $\hpm$ with up to two strips or three $\hpm$ with up to one strip, representing the decay modes mentioned above, are constructed in parallel. In case there exist multiple possibilities for choosing the $\hpm$ among the jet constituents or for choosing the strips among the set of strips reconstructed in the previous step, the HPS algorithm constructs all possible combinations among the $6$ highest $\pT$ $\hpm$ and the $6$ highest $\pT$ strips. The restriction to the $6$ highest $\pT$ objects is imposed to reduce the computational complexity of the algorithm.

The constructed $\tauh$ candidates are subject to preselection criteria, which demand the sum of $\hpm$ charges to be equal to $\pm 1$, all $\hpm$ and strips to be within a signal cone of radius $\Delta R = 3.0/(\pT \, \GeV^{-1})$ (limited to a minimum of $0.05$ and a maximum of $0.10$), and the mass of the $\tauh$ candidate to be within a certain mass window~\cite{CMS:2015pac}. 
The signal cone is centred on the momentum vector of the $\tauh$ candidate. 
The distance between the $\tauh$ candidate and the $\hpm$ or strip is computed according to Eq.~(\ref{eq:deltaR}).

The four-vector of the $\tauh$ candidate is computed by summing the four-vectors of its constituent $\hpm$ and strips. The energy of electrons and positrons that are considered as $\hpm$ is adjusted such that their mass matches the $\Pgppm$ meson mass when summing the four-vectors. In order to avoid double-counting of $\Pe$ in case they are considered as $\hpm$ and have been clustered into strips, the $\Pe$ considered as $\hpm$ are removed from the strips and the strip momentum is recomputed. In case there remains no particle in the strip after removing these $\Pe$, the $\tauh$ candidate is discarded. 

In case multiple $\tauh$ candidates pass the preselection criteria, the $\tauh$ candidate of highest $\pT$ is retained and all other $\tauh$ candidates corresponding to the same jet are discarded.

\subsection*{DeepTau}

The DeepTau algorithm is published in Ref.~\cite{CMS:2022prd}.
The purpose of the algorithm is to identify the $\tauh$ that are reconstructed by the HPS algorithm as described in the previous section. 

\begin{figure*}[ht!]
    \centering
    \ifx\ver\verPreprint
        \includegraphics[width=1.00\textwidth]{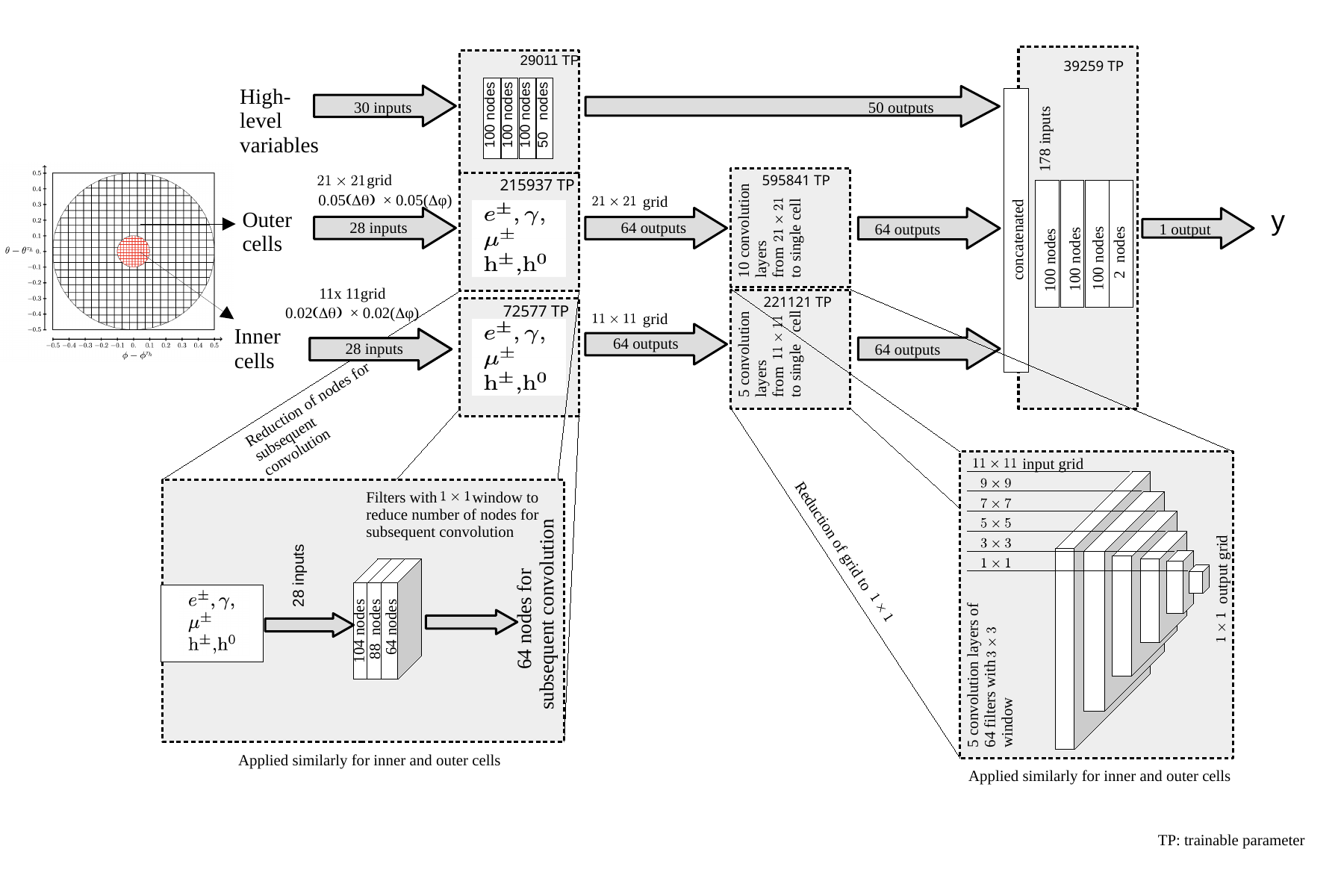}
    \fi
    \ifx\ver\verPAPER
        \includegraphics[width=0.95\textwidth]{Figs/DNN.pdf}
    \fi
    \caption{
      The DNN architecture of the DeepTau algorithm.
      The numbers of trainable parameters (TP) for different components of the network are given in the figure.
    }
    \label{fig:DeepTau_architecture}
\end{figure*}

The DNN architecture of the DeepTau algorithm is illustrated in Fig.~\ref{fig:DeepTau_architecture}.
It is composed of three subnetworks. Two subnetworks process the information about
individual particles near the $\tauh$, while the third subnetwork processes a set of high-level features of the $\tauh$.

The first and second subnetworks have similar structure.
The information about location, type, and other properties of the particles near the $\tauh$ are discretised in two grids in the $\theta$--$\phi$ plane, an inner grid of $11 \times 11$ cells of size $0.02 \times 0.02$ and an outer grid of $21 \times 21$ cells of size $0.05 \times 0.05$.
The finer segmentation of the inner grid reflects the fact that the particles produced in hadronic $\Pgt$ decays are typically highly collimated (\cf Fig.~\ref{fig:discrObservables}) and furthermore helps to resolve the dense core of high-energetic quark and gluon jets~\cite{CMS:2022prd}.
The inner and outer grids are centred on the direction of the $\tauh$. The grids are populated by iterating over all particles reconstructed in the event and computing their distance in $\theta$ and $\phi$ with respect to the $\tauh$ direction. Particles falling into the same cell are sorted in the order of decreasing $\pT$ regardless of their type.
The location, type, and other properties of the two particles of highest $\pT$ are concatenated to a vector of size $28$. The particle properties used to build this vector are given in Table~\ref{tab:DeepTau_lowlevelinputs}. We use $14$ properties per particle.
Cells in the outer grid that overlap with the inner grid are skipped when populating the grids,
in order to avoid redundancy of information between the two grids~\footnote{
The CMS implementation of the DeepTau algorithm uses a few more particle properties as inputs, which are not available in our simulation.
The missing inputs concern mainly detector-level observables, which are used to improve the separation of $\tauh$ from electrons and muons in CMS.
We expect these observables to have little effect on the performance in separating $\tauh$ from quark and gluon jets.
Our implementation of the DeepTau algorithm differs from the implementation in CMS in three further aspects:
First, we compute distances between particles in the $\theta$--$\phi$ plane,
while CMS computes the distances in the $\eta$--$\phi$ plane, where the symbol $\eta = -\ln\left(\tan\frac{\theta}{2}\right)$ denotes the pseudo-rapidity.
Second, CMS builds three separate inner grids and three separate outer grids for particles of types $\Pe/\Pgg$, $\Pgm$, and $\hpm/\hzero$
and for each type considers only the particle of highest $\pT$ when building the vector of particle properties.
We find that we get better performance if we use a single inner and a single outer grid for particles of any type
and instead consider up to two particles (of highest $\pT$) per cell.
Third, CMS does not remove the overlap between the inner and outer grids.
We find that, besides avoiding redundancy of information, removing this overlap also improves the performance (by a small amount).
We have adjusted the size of subsequent layers in the network according to these differences.}.
The features in each cell of the inner and outer grids are preprocessed by four fully-connected layers of size $104$, $88$, $64$. The preprocessed information is then passed through a stack of $5$ convolutional layers for the inner grid and $10$ for the outer grid.
Each convolutional layer uses $64$ filters and a kernel of size $3 \times 3$.
Because the convolutional layers use no padding, the size of the grid decreases by two units in $\theta$ and two units in $\phi$ per convolutional layer. The output of the two subnetworks for the inner and outer grids are two single cells that each hold a vector of size $64$, resulting from the application of the $64$ filters.

The high-level features processed by third subnetwork are given in Table~\ref{tab:DeepTau_highlevelinputs}. We use $30$ high-level features in total.
The variable $I_{\textrm{ch3}}$ ($I_{\textrm{ch5}}$) refers to the isolation of the $\tauh$ with respect to charged particles ($\Pe$, $\Pgm$, $\hpm$),
which are within an ``isolation cone'' of size $\Delta R = 0.3$ ($0.5$) and were not used to build the $\tauh$ object by the HPS algorithm:

\begin{equation}\label{eq:Itau}
    I_{\textrm{ch}} = \sum_{\textrm{charged}} \, \pT \, .
\end{equation}

The variables $I_{\Pgg3}$ and $I_{\Pgg5}$ ($I_{\textrm{nh3}}$ and $I_{\textrm{nh5}}$) are computed similarly by summing the $\pT$ of all particles of type $\Pgg$ ($\hzero$)
within the isolation cone.
The variable $\langle r^{\Pgg} \rangle$ is computed using Eq.~(\ref{eq:jetRadius}), with the sum extending over all $\tauh$ constituents of type $\Pgg$
and distance $\Delta R$ computed between the $\tauh$ constituent and the direction of the $\tauh$.
The variables $\langle r_{\theta}^{\Pgg} \rangle$ and $\langle r_{\phi}^{\Pgg} \rangle$ are computed analogously, but taking only differences in either $\theta$ or $\phi$ into account.
The high-level features are processed by four fully-connected layers, with a size of $100$ for the first three layers and a size of $50$ for the fourth layer.

The output of the fully-connected layers that processed the high-level features is concatenated with the outputs of the two subnetworks for the inner and outer grids.
The resulting vector of size $178$ is passed through three fully-connected layers of size $100$.
The discriminant $\Dtau$ of the network is computed by a layer of size $1$.
The softmax activation function~\cite{10.1007/978-3-642-76153-9_28} is used for this last layer, while all other fully-connected layers and the convolutional layers use the PReLU activation function~\cite{he2015delving}.

\begin{table*}[ht!]
    \centering
    \begin{tabular}{c|l}
        Variable & Description \\
        \hline
        $\pT^{i} / \pT^{\Pgt}$ & $\pT$ of particle $i$ in relation to $\tauh$ \\
        $\Delta\theta$, $\Delta\phi$ & distance between particle $i$ and $\tauh$ in polar and azimuthal direction \\
        $M_{i}$ & mass of particle $i$ \\
        charge & electric charge of particle $i$ \\
        $\dxy$, $\sigmaxy$ & transverse impact parameter and its uncertainty \\
        $\dz$, $\sigmaz$ & longitudinal impact parameter and its uncertainty \\
        $\indicator_{\Pe}$,$\indicator_{\Pgm}$,$\indicator_{\Pgg}$,$\indicator_{\ch}$,$\indicator_{\nh}$ & type of particle $\Pe$, $\Pgm$, $\Pgg$, $\hpm$, $\hzero$ in one-hot-encoded format \\
    \end{tabular}
    \caption{
      Properties of particles used as input to the two subnetworks that process the inner and outer grids in the DeepTau algorithm.
      The variables $\Delta\theta$ and $\Delta\phi$ are defined as $\Delta\theta = \theta_{i} - \theta_{\Pgt}$ and $\Delta\phi = \phi_{i} - \phi_{\Pgt}$, where the subscript $\Pgt$ refers to the direction of the $\tauh$ and $i$ to that of a particle in the inner or outer grid.
      The variables $\dxy$, $\sigmaxy$, $\dz$, and $\sigmaz$ are computed as described in Ref.~\cite{christian_veelken_2023_8113344}. They are set to zero if the particle is of type $\Pgg$ or $\hzero$.
    }
    \label{tab:DeepTau_lowlevelinputs}
\end{table*}

\begin{table*}[ht!]
    \centering
    \begin{tabular}{c|l}
        Variable & Description \\
        \hline
        $\pT^{\Pgt}$, $\theta_{\Pgt}$, $\phi_{\Pgt}$, $M_{\Pgt}$ & $\pT$, $\theta$, $\phi$, and mass of $\tauh$ \\ 
        charge & $\tauh$ charge (equal to sum of $\hpm$ charges) \\
        $I_{\textrm{ch5}}$, $I_{\Pgg5}$, $I_{\textrm{nh}5}$ & isolation of the $\tauh$ with respect to charged particles, $\Pgg$, and $\hzero$, \\
         & computed for an isolation cone of size $\Delta R = 0.5$ \\
        $I_{\textrm{ch3}}$, $I_{\Pgg3}$, $I_{\textrm{nh}3}$ & isolation of the $\tauh$ with respect to charged particles, $\Pgg$, and $\hzero$, \\
         & computed for an isolation cone of size $\Delta R = 0.3$ \\
        $N_{\ch}$, $N_{\Pgg}$ & multiplicity of $\tauh$ constituents of type $\hpm$ and $\Pgg$ \\
        $\max(\pT^{\ch})$ & maximum $\pT$ among $\tauh$ constituent of type $\hpm$ \\
        $\sum \pT^{\Pgg} /\pT^{\Pgt}$ & $\pT$-sum of $\tauh$ constituents of type $\Pgg$, divided by $\pT$ of $\tauh$ \\
        $\pT^{\Pgg\textrm{,out}}$ & $\pT$-sum of $\Pgg$, which are in strips, but outside signal cone \\
        $\langle r^{\Pgg} \rangle$, $\langle r_{\theta}^{\Pgg} \rangle$, $\langle r_{\phi}^{\Pgg} \rangle$ & $\pT$-weighted distances in $\Delta R$, $\theta$, and $\phi$ between $\tauh$ and $\Pgg$ \\
        $\langle r^{\Pgg\textrm{,out}} \rangle$ & $\pT$-weighted distance in $\Delta R$ between $\tauh$ and $\Pgg$, \\
         & which are in strips, but outside signal cone \\
        $N_{\Pe}^{\ig}$, $N_{\Pgm}^{\ig}$, $N_{\Pgg}^{\ig}$, $N_{\ch}^{\ig}$, $N_{\nh}^{\ig}$ & number of particles of type $\Pe$, $\Pgm$, $\Pgg$, $\hpm$, $\hzero$ in inner grid \\
        $N_{\Pe}^{\og}$, $N_{\Pgm}^{\og}$, $N_{\Pgg}^{\og}$, $N_{\ch}^{\og}$, $N_{\nh}^{\og}$ & number of particles of type $\Pe$, $\Pgm$, $\Pgg$, $\hpm$, $\hzero$ in outer grid \\
    \end{tabular}
    \caption{
      High-level features used as input to the DeepTau algorithm.
    }
    \label{tab:DeepTau_highlevelinputs}
\end{table*}

The network is implemented in \textsc{PyTorch}~\cite{paszke2019pytorch}.
It has $1.6 \cdot 10^{6}$ trainable parameters, which are trained in batches of $500$ jets for a maximum of $200$ epochs, using the AdamW~\cite{loshchilov2019decoupled} optimizer.
A fixed learning rate of $10^{-4}$ is used throughout the training.
For the loss function, we use the focal loss~\cite{Lin:2017fqe},
with a value $\gamma = 2$ for the focusing parameter $\gamma$.
The robustness of the training is increased by applying layer normalisation~\cite{ba2016layer} to the inputs of each fully-connected and convolutional layer.
The loss on the validation dataset is monitored throughout the training and the model with the minimum validation loss is retained for further study.

\begin{figure}[htb]
    \centering
    \ifx\ver\verPreprint
        \includegraphics[width=0.52\textwidth]{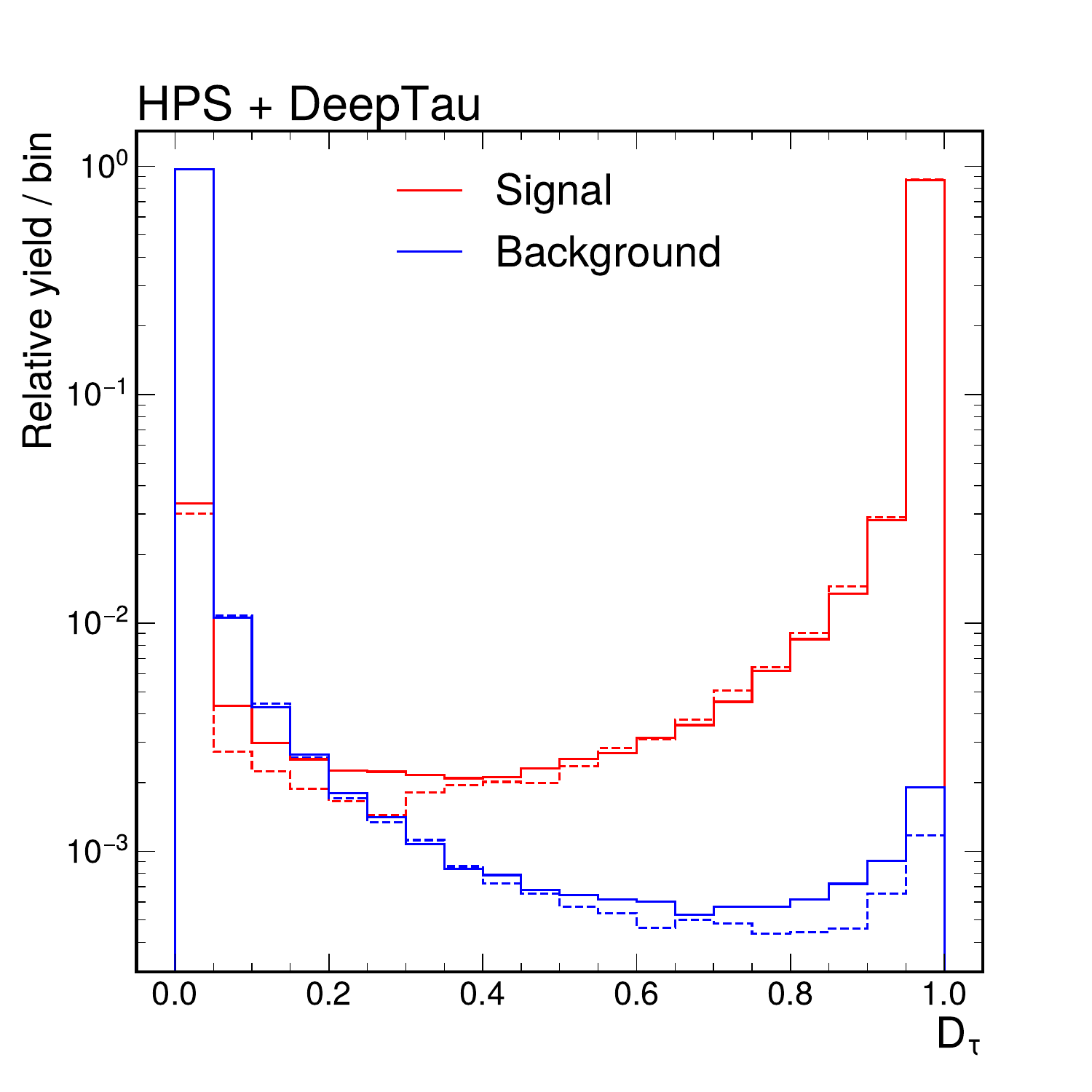}
    \fi
    \ifx\ver\verPAPER
        \includegraphics[width=0.95\columnwidth]{Figs/tauClassifiers/HPS_DeepTau_tauClassifier.pdf}
    \fi
    \caption{
     Distribution in the discriminant $\Dtau$ for the DeepTau algorithm.
     The solid curves refer to the test dataset and the dashed curves to the training dataset.
    }
    \label{fig:tauClassifier_Deeptau}
\end{figure}

We show the distribution in the discriminant $\Dtau$ computed by the DeepTau algorithm in Fig.~\ref{fig:tauClassifier_Deeptau},
separately for the training and test datasets.

\ifx\ver\verPreprint
    \clearpage
\fi
\bibliography{ml-tau}

\end{document}